\documentclass[structabstract]{aa}  
\usepackage{graphicx}
\usepackage{txfonts}
\begin{document}
   \title{Extremely faint high proper motion objects from SDSS stripe 82\thanks{Based on observations with VLT/FORS1 at the European Southern Observatory (ESO program 078.D-0595)}}

   \subtitle{Optical classification spectroscopy of about 40 new objects}

   \author{R.-D. Scholz\inst{1}\and
           J. Storm\inst{1}\and
           G.~R. Knapp\inst{2}\and
           H. Zinnecker\inst{1}
          }
   \institute{Astrophysikalisches Institut Potsdam, 
              An der Sternwarte 16, 14482 Potsdam, Germany\\
              \email{rdscholz@aip.de, jstorm@aip.de, hzinnecker@aip.de}
          \and
              Department of Astrophysical Sciences, Princeton University, 
              Peyton Hall, Princeton, NJ 08544, USA\\
              \email{gk@astro.princeton.edu}
             }

   \date{Received 29 September 2008; accepted ...}


  \abstract
   {}
   {By pushing the magnitude limit of high proper motion surveys beyond
    the limit of photographic Schmidt plates we aim at the discovery
    of nearby and very fast low-luminosity objects of different classes:
    cool white dwarfs (CWDs), cool subdwarfs (sd), and very low-mass stars 
    and brown dwarfs at the very faint end of the main sequence (MS).
   }
   {The deep multi-epoch Sloan Digital Sky Survey data in a 275 square degrees
    area along the celestial equator (SDSS stripe 82) allow us to search for 
    extremely faint ($i>21$) objects with proper motions larger than 0.14~arcsec/yr.
    A reduced proper motion diagram $H_z/(i-z)$ clearly reveals three sequences 
    (MS, sd, CWD) where our faintest candidates are representative of the 
    still poorly known bottom of each sequence.
    We classify 38 newly detected objects with low-resolution optical 
    spectroscopy using FORS1 @ ESO VLT. Together with our targets we
    observe six known L dwarfs in stripe 82, four (ultra)cool sd and one CWD
    as comparison objects. Distances and tangential velocities are estimated
    using known spectral type/absolute magnitude relations.
   }
   {All 22 previously known L dwarfs (and a few of the T dwarfs) in stripe 82
    have been detected in our high proper motion survey. However, 11 of the known
    L dwarfs have smaller proper motions (0.01$<$$\mu$$<$0.14~arcsec/yr).
    Although stripe 82 was already one of the best investigated sky regions with
    respect to L and T dwarfs, we are able to classify 13 new L dwarfs. Two
    previously known L dwarfs have been reclassified by us. We have also found 
    eight new M7.5-M9.5 dwarfs. The four new CWDs discovered by us
    are about 1-2 mag fainter than those previously detected in SDSS data.
    All new L-type, late-M and CWD objects show thick disk and halo kinematics. 
{Since our high-velocity late-M and L dwarfs do not show indications
of low metallicity in their spectra, we conclude that there may be a
population of ultracool halo objects with normal metallicities.
There are 13 objects, mostly with uncertain proper motions, which we
    initially classified as mid-M dwarfs. Among them we have found 9 with an 
    alternative subdwarf classification (sdM7 or earlier types),
    whereas we have not found any new spectra resembling the
    known ultracool ($>$sdM7) subdwarfs.
    Some M subdwarf candidates have been classified based on 
spectral indices with large uncertainties.}
    We failed to detect new nearby ($d<50$~pc)
    L dwarfs, probably because the SDSS stripe 82 area was already well-investigated before. 
    With our survey we have demonstrated a higher efficiency in finding
    Galactic halo CWDs than previous searches. 
    The space density of halo CWDs is according to our results 
about 1.5-3.0 $\times$ 10$^{-5}$~pc$^{-3}$.
   }
   {}


   \maketitle
%

\section{Introduction}

Deep sky surveys are not only important for the study of the most distant
galaxies in the Universe but also for near-field stellar statistics and 
physics. 
It has been realised for a long time that the immediate Solar neighbourhood
is the only region of a galaxy where there is a real chance for a complete
registration of its stellar and sub-stellar content reaching to the 
lowest-luminosity objects. A complete
volume-limited sample provides the groundwork for the knowledge of the local
luminosity function, the mass function of field stars and brown dwarfs,
and the Galactic
star formation history (Reid et al.~\cite{reid02};
Gizis et al.~\cite{gizis02}; Cruz et al.~\cite{cruz03}).
Moreover, the nearest representatives of a given class
of objects can be considered as benchmark sources allowing detailed follow-up
studies, including the search for possible planetary companions.

However, our knowledge of the Solar neighbourhood is still remarkably 
incomplete: Within a 25~pc horizon more than 60\% of the stars 
(Henry et al.~\cite{henry02})
and probably more than 90\% of the brown dwarfs remain to be
discovered. The Research Consortium on Nearby Stars (RECONS)
has documented the recent progress achieved for the 10~pc 
sample\footnote{http://www.chara.gsu.edu/RECONS/census.posted.htm}.

It should be noted that this sample is also not complete 
since it neglects recently discovered sources likely within 10~pc but 
without accurately measured parallaxes. It
is defined as all known objects with trigonometric parallaxes of 100~mas
or more, and an error in that parallax of less 
than 10~mas. An increase of about 20\% has been achieved since 2000,
so that the census has reached about 350 objects. About 240 of them
are M dwarfs, which also contributed to the increase with the largest
number (+40). The only other object classes with improved statistics
are the L/T dwarfs and the planetary companions around nearby stars.
The number of white dwarfs (18) is stagnating, but there are ongoing 
efforts to find missing nearby white dwarfs (Subasavage et 
al.~\cite{subasavage04}; \cite{subasavage08}). 
Subdwarfs represent the smallest group
with only 4 objects in the 10~pc sample (Henry et al.~\cite{henry06}),
which all have been known for a long time, and which all have extremely large
proper motions (3.7$<$$\mu$$<$8.7~arcsec/yr).

Before time-consuming trigonometric parallax measurements can be started,
one needs to know good nearby candidates of low-luminosity stars and
substellar objects. Most nearby stars
have been first detected in high proper motion (HPM) surveys 
(e.g. Luyten~\cite{luyten79a,luyten79b}; Luyten \& Hughes~\cite{luyten80}). 
HPM objects represent a mixture of {\it very nearby}
neighbours of the Sun in the local population of the Galactic thin
disk, and {\it very fast} representatives of the Galactic thick disk
and halo, just passing through the neighbourhood.
Although halo stars are relatively rare
when compared to the number of disk stars in a local volume-limited
sample, they are over-represented in HPM samples.
The main observational material
used in Luyten's HPM surveys were photographic Schmidt plates.  
Due to their full sky coverage in different colours and over long time 
intervals these archival data continue to play an important role in the
search for nearby candidates. In particular, the deeper and partly redder
($I$-band) second epoch Schmidt survey plates have by far not yet been fully
exploited to their limiting magnitudes.

The SuperCOSMOS Sky Surveys (SSS) represent one of the best available 
photographic archives suitable for the identification of hitherto
unknown faint HPM stars, since they provide multi-colour catalogue data with
cross-matching between the different passbands and epochs as well as 
images from the individual digitised photographic plates taken with
three different Schmidt 
telescopes (Hambly et al.~\cite{hambly01a,hambly01b,hambly01c}). 
The comparison of optical SSS data with near-infrared Two Micron All 
Sky Survey (2MASS; Skrutskie et al.~\cite{skrutskie06}) 
led to the discovery of new faint HPM objects with $\mu>$ 1 arcsec/yr 
including some of the nearest brown dwarfs (Scholz et al.~\cite{scholz03};
Kendall et al.~\cite{kendall07}; Folkes et al.~\cite{folkes07}),
cool subdwarfs (Scholz et al.~\cite{scholz04a,scholz04b}), late-type M
dwarfs (Pokorny et al.~\cite{pokorny03}; 
Hambly et al.~\cite{hambly04}), and cool white dwarfs 
(Scholz et al.~\cite{scholz02}; Hambly et al.~\cite{hambly04}).

New near-infrared sky surveys like the 2MASS and the DEep Near-Infrared
Survey (DENIS; Epchtein et al.~\cite{epchtein97}) and deep multi-colour 
optical surveys like the Sloan Digital
Sky Survey (SDSS; York et al.~\cite{york00}) 
uncovered large numbers of objects cooler than the
latest-type M dwarfs. New spectral classes, L (2500~K$>$$T_{eff}$$>$1500~K)
and T (1500~K$>$$T_{eff}$$>$700~K), describing these
ultracool objects, had to be defined less than ten years ago
(Mart\'{\i}n et al.~\cite{martin99};
Kirkpatrick et al.~\cite{kirkpatrick99}; Burgasser et al.~\cite{burgasser02}
Geballe et al.~\cite{geballe02}).
Since then almost 650 L and T dwarfs have been discovered, mostly in 2MASS
($\approx$50\%) and SDSS ($\approx$30\%)
(see Gelino et al.~\cite{gelino08}, and references therein).
About 50\% of the L dwarfs and all T dwarfs in the
Solar neigbourhood are brown dwarfs, whereas among late-M dwarfs only a
few are sub-stellar objects (see review by Kirkpatrick~\cite{kirkpatrick05}).
Note that one of the latter is LP\,944-20 (Tinney~\cite{tinney98}), the only
isolated brown dwarf catalogued already by Luyten~(\cite{luyten79b};
Luyten \& Hughes~\cite{luyten80}) as an HPM object, long
before the first free-floating brown dwarfs were discovered from a new HPM 
survey by Ruiz et al.~(\cite{ruiz97}) and from DENIS
(Delfosse et al.~\cite{delfosse97}). 

Cool, low-mass subdwarfs are interesting as tracers of Galactic structure
and chemical evolution. 
They are metal-poor and typically exhibit halo
kinematics (hence they show up preferentially in HPM surveys). 
The current classification scheme of cool subdwarfs ends
at sdM7 (Gizis~\cite{gizis97}). However, more and more cooler subdwarfs
are being discovered in recent years (e.g. L{\'e}pine et al.~\cite{lepine03};
Burgasser et al.~\cite{burgasser03}; Scholz et al.~\cite{scholz04a,scholz04b};
Burgasser \& Kirkpatrick~\cite{burgasser06b}). A systematic search in the 
SDSS spectroscopic database has uncovered 23 new subdwarfs with spectral
sub-types of M7 or later and more than doubled the number of known ultracool 
subdwarfs (L{\'e}pine \& Scholz~\cite{lepine08}). 
There is some evidence that
these so-called ultracool subdwarfs are somewhat hotter
compared to normal dwarfs of the same spectral sub-type. 
Burgasser et al.~(\cite{burgasser07}) derived 3100~K$>$$T_{eff}$$>$2400~K 
for subdwarfs with spectral sub-types between M7.5 and L4 from 
comparison with model spectra.
An extension of the classification scheme for the ultracool subdwarfs,
including sub-stellar subdwarfs, is now 
developing, where 
the spectral features 
indicating subtle changes in their  physical properties (metallicity,
surface temperature/clouds, gravity) are still under debate 
(Kirkpatrick~\cite{kirkpatrick05}; Gizis \& Harvin~\cite{gizis06};
L{\'e}pine et al.~\cite{lepine07}; Burgasser et al.~\cite{burgasser07}).

A third class of cool objects, completely different in their evolutionary 
stage and physical parameters from low-mass 
stars/brown dwarfs and cool subdwarfs, but also hard to detect due to their
faintness, are the cool white dwarfs (CWDs). CWDs have long been 
considered as important objects for a number of reasons, such as
determining the age of the Galactic disk (Leggett et al.~\cite{leggett98}) 
and as possible components of the Galactic dark 
matter halo (Ibata et al.~\cite{ibata00}; 
Oppenheimer et al.~\cite{oppenheimer01};
M{\'e}ndez~\cite{mendez02}; Flynn et al.~\cite{flynn03};
Salim et al.~\cite{salim04}; Reid~\cite{reid05}). Among the coolest known
CWDs ($T_{eff}$$<$4500~K) in the Solar neighbourhood
there are many objects already catalogued in the LHS 
(Luyten~\cite{luyten79a}) or discovered as HPM objects from photographic plates
(Hambly et al.~\cite{hambly97}; Harris et al.~\cite{harris99};
Ibata et al~\cite{ibata00}; Scholz et al.~\cite{scholz00}; 
Monet et al.~\cite{monet00}; Scholz et al.~\cite{scholz02};
L{\'e}pine et al.~\cite{lepine05}; Carollo et al.~\cite{carollo06};
Rowell et al.~\cite{rowell08}).
Other CWDs have been found in the 
SDSS (Harris et al.~\cite{harris01}; Gates et al.~\cite{gates04}; 
Kilic et al.~\cite{kilic06}; Vidrih et al.~\cite{vidrih07};
Harris et al.~\cite{harris08}).
Hall et al.~(\cite{hall08}) have recently found a nearby halo CWD
with a large proper motion detected by combining the photographic USNO-B1.0
catalogue (Monet et al.~\cite{monet03}) with the SDSS 
(Munn et al.~\cite{munn04}).
The space density of CWDs in the Galactic halo, possibly
responsible for observed microlensing events, is still
under debate (see e.g. Torres et al.~\cite{torres08}). Reid et 
al.~(\cite{reid01} had shown that the halo CWDs detected by
Oppenheimer et al.~(\cite{oppenheimer01}) in a HPM survey
based on photographic Schmidt plates are in fact members of the
thick disk. We conclude that a deeper HPM survey could either 
detect cooler objects or uncover the CWDs with even higher 
velocities, i.e. the real halo members.

As we have shown above,
extending HPM surveys to fainter magnitude limits has led to the
discovery of many benchmark sources of new classes of cool objects.
Several new large-area HPM surveys relying {\it only} on modern deep
optical (SDSS) and/or near-infrared (2MASS and SIMP) data have 
been successfully started
(Artigau et al.~\cite{artigau06}; 
Looper et al.~\cite{looper07}; Vidrih et al.~\cite{vidrih07};
Bramich et al.~\cite{bramich08};
Metchev et al.~\cite{metchev08}). The UKIRT Infrared Deep Sky Survey 
(UKIDSS; Lawrence et al.~\cite{lawrence07}) will eventually
provide its own second epoch data suitable for HPM searches for the
coolest brown dwarfs. The large-area survey of UKIDSS
overlaps with SDSS, hence these different epoch data enable already
deep HPM searches.

In this paper we present the first results of one of 
the deepest optical HPM searches done so far over a relatively 
large sky area.  It is based on multi-epoch SDSS data in 
an equatorial stripe (called stripe 82) of about 275 square degrees. 
In Sect.~\ref{sec2} we describe our HPM survey and the selection 
of targets for follow-up observations. Sect.~\ref{sec3} deals with the
spectroscopic classification of our extremely faint HPM objects. In
Sect.~\ref{sec4} we discuss their distance estimates and the resulting 
kinematics. In Sect.~\ref{sec5} we draw some conclusions and give an outlook
on further work needed.

   \begin{figure*}[!]
   \centering
   \includegraphics[angle=270,width=19cm]{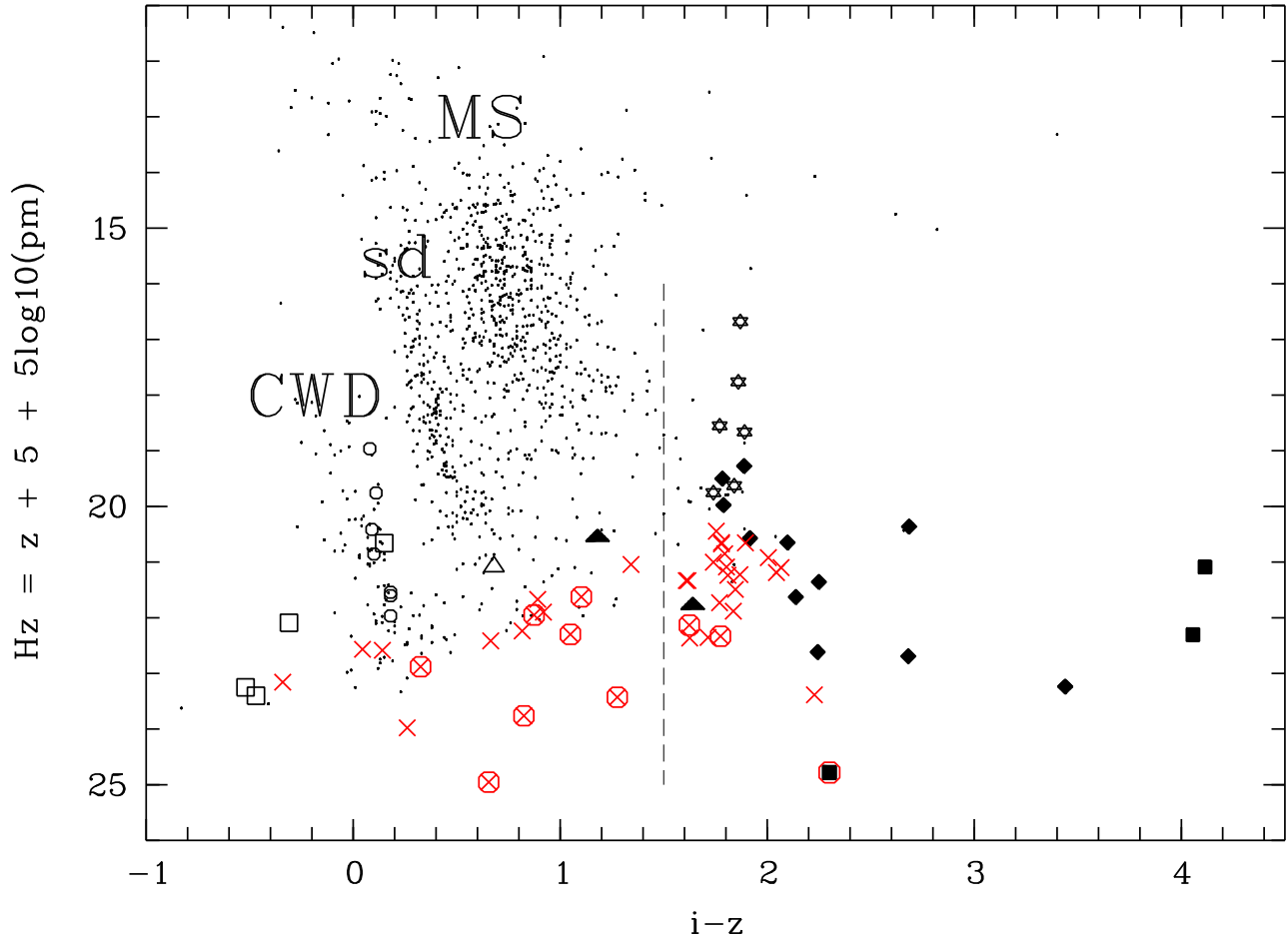}
   \caption{Reduced proper motion diagram for objects with
proper motions exceeding 0.14~arcsec/yr: all initially found
objects after step 1) of our HPM survey (dots) with three
separate sequences (MS, sd, CWD) marked,
known L dwarfs (filled lozenges) and T dwarfs (filled squares) falling
in the S82 survey area, M9 dwarfs
(stars), L-type subdwarfs (filled triangles), a late-M extreme (esdM6.5)
subdwarf (open triangle), CWDs (open circles and squares), and the confirmed
faintest HPM objects after step 2) and 3) of our survey (crosses). Symbols 
overplotted by large open circles mark objects with uncertain proper motions 
(see Tab.~\ref{tabnewhpm} and Tab.~\ref{tabLTpm}).
               }
              \label{fig1}%
    \end{figure*}
%

%

\section{HPM survey and target selection}
\label{sec2}

\subsection{SDSS stripe 82}
\label{s82hpm}

The SDSS consists of an imaging and an spectroscopic survey over more than
25\% of the sky (York et al.~\cite{york00}).
SDSS imaging data are taken simultaneously in the $u, g, r, i, z$ bands
(Fukugita et al.~\cite{fukugita96}; Gunn et al.~\cite{gunn98}; 
Smith et al.~\cite{smith02}; Gunn et al.~\cite{gunn06}) in drift-scan
mode with a chip arrangement leading to a filled stripe of about 2.5 degrees
width after two slightly overlapping scans (north and south strips). The
data are pipeline-reduced providing accurate astrometry 
(Pier et al.~\cite{pier03}) and photometry 
(Lupton et al.~\cite{lupton99,lupton01}; Hogg et al.~\cite{hogg01}; 
Stoughton etal.~\cite{stoughton02}; Ivezic et al.~\cite{ivezic04};
Tucker et al.~\cite{tucker06}; Padmanabhan et al.~\cite{padmanabhan08}). 
SDSS stripe 82 (hereafter S82)
is the equatorial stripe ($\delta$ from $-1.25$ to $+1.25$ degrees) spanning
8 hours in right ascension ($\alpha$ from $20^h$ to $4^h$). S82 north and south
strips have been observed repeatedly beginning in 1998. 
Since an imaging run may have covered all or part of a strip,
about 250 square degrees in S82 were already
covered by more than 20 SDSS imaging epochs until the end of 2004. This
was the status in the SDSS database available at Princeton (Finkbeiner
et al.~\cite{finkbeiner04}), which we considered as an excellent
starting point for our search for faint moving objects. 

Note that in our study we have not made use of the light and motion
catalogues in S82 recently published by Bramich et al.~(\cite{bramich08}).
Compared to the CWD candidates extracted from the
latter by Vidrih et al.~(\cite{vidrih07}), our HPM targets selected for
VLT spectroscopy (see Sect.~\ref{rpmd}) are generally fainter by about
one magnitude. Our S82 HPM survey was initiated in November 2005, about
one year before a similar project by Bramich, Vidrih and coworkers was
started. From the beginning of our independent S82 survey we were more 
interested in exploring the faintest possible HPM detections rather than 
in studying the whole HPM sample or presenting a complete catalogue.
S82 data taken after the end of 2004, in particular the SDSS-II Supernova
Survey (Frieman et al.~\cite{frieman08}), were not included in our study.

We have carried out our HPM survey primarily based
on the SDSS multi-epoch imaging data in S82 taken between September 1998 and 
December 2004, but also using available 2MASS data for verification.
This survey goes more than two magnitudes deeper compared to the
SSS based Southern sky HPM survey, which led to the discovery of the nearest 
brown dwarf, $\varepsilon$ Indi B (Scholz et al.~\cite{scholz03}),
and also in comparison to the new complete Northern sky HPM survey by
L{\'e}pine \& Shara~(\cite{lepine05b}) based on photographic Schmidt plates.
The high accuracy of SDSS single epoch astrometry
has been demonstrated by Pier et al.~(\cite{pier03}).
These data allow us already
to determine proper motions, e.g. of faint brown dwarfs, if several 
epochs are available over a relatively short time interval of only few 
years (Finkbeiner et al.~\cite{finkbeiner04}). 

Our HPM survey in S82 was carried out in stages:
{\bf 1)} For the first step we selected 
(due to restrictions in the computing power initially available
for our task)
a limited number of overlapping SDSS runs, well distributed in time
(separated by about 6 months)
over the full baseline of about six years. Tools described in
Finkbeiner et al.~(\cite{finkbeiner04}) were used for extracting 
HPM candidates out of the Princeton SDSS database. A minimum of 5 epoch
measurements was required for determining initial proper motions, typically
9 epochs over 3 to 6 years were used. The resulting survey area was
about 275 square degrees. {\bf 2)} All objects with initial proper motions 
above 0.2 arcsec/yr were then investigated in more detail in a second step,
involving all available SDSS epochs. For objects significantly brighter than 
the SDSS magnitude limits (a 95\% detection repeatability for point sources
is provided at $u=22.0, g=22.2, r=22.2, i=21.3, z=20.5$) 
there were about 15 to 35 epochs depending on the
different overlap status within S82. However, objects at the SDSS magnitude
limits, in which we were especially interested, did not appear on all 
available SDSS images. The 
proper motion fit was then checked visually, including a comparison of
available 2MASS positions with the predicted position from the SDSS proper
motion, to ensure that we select only real HPM
objects. For the red objects ($i-z > 1$) we were able
to extend the visual checking to a lower proper motion limit of about
0.05~arcsec/yr. 
This was possible since the number of these candidates was 
not too large (less than a few thousand)
{\bf 3)} Finally, multi-epoch SDSS finding charts were 
inspected and data points with strongly deviating photometry were excluded
from the proper motion fit in order to exclude doubtful HPM candidates.
In particular, we excluded objects with matching problems between different
epochs (close neighbors) and 
those of the
faint objects, for which a large proper motion was
caused by a 
\textit{different position at only one epoch} 
(may be in fact two different
objects at the SDSS magnitude limit).  

%
\begin{table*}
\caption{New faint HPM stars detected from SDSS S82 data}
\label{tabnewhpm}
\centering
\begin{tabular}{l c c c c c r r c}
\hline\hline
Name from SDSS     & $\alpha$ (J2000) & $\delta$ (J2000) & Epoch  & mean $i$ & mean $z$ &$\mu_{\alpha}\cos{\delta}$ & $\mu_{\delta}$ & $N_{ep}$ \\
DR6 (SDSS~J...)    & [\degr]          & [\degr]          & [year] & [mag]    & [mag]    & [mas/yr]                  &  [mas/yr]      & \\
(1)                & (2)              & (3)              & (4)    & (5)      & (6)      & (7)                       & (8)            & (9) \\
\hline
002138.86$+$002605.8 & 005.411918 & $+$00.434957 & 2002.6785 & 21.744$\pm$0.045 & 19.934$\pm$0.024 & $+$161.4$\pm$12.6 & $-$84.6$\pm$09.8 & 13 \\ 
002800.37$+$010252.0 & 007.001579 & $+$01.047793 & 2002.6785 & 21.637$\pm$0.034 & 19.898$\pm$0.022 & $+$164.4$\pm$10.6 & $-$25.1$\pm$15.6 & 32 \\ 
003435.32$+$004633.9 & 008.647194 & $+$00.776103 & 2001.7887 & 21.418$\pm$0.042 & 19.664$\pm$0.021 & $+$136.9$\pm$10.9 & $+$42.2$\pm$07.5 & 22 \\ 
003550.48$+$004631.3 & 008.960348 & $+$00.775378 & 2001.7887 & 21.923$\pm$0.065 & 20.314$\pm$0.042 & $+$152.2$\pm$14.4 & $-$48.9$\pm$13.8 & 23 \\ 
005349.11$-$011354.4 & 013.454634 & $-$01.231797 & 2002.6786 & 21.638$\pm$0.052 & 19.860$\pm$0.031 & $+$144.7$\pm$24.2 & $+$15.0$\pm$12.0 & 14 \\ 
005636.22$-$011131.9 & 014.150936 & $-$01.192200 & 2002.6786 & 22.118$\pm$0.082 & 20.074$\pm$0.062 & $-$1.0$\pm$16.6 & $-$167.0$\pm$31.1 & 13 \\ 
005837.48$+$004435.6 & 014.656173 & $+$00.743229 & 2001.7888 & 21.356$\pm$0.046 & 19.731$\pm$0.026 & $+$311.6$\pm$08.6 & $-$126.0$\pm$12.0 & 23 \\ 
010959.91$-$010156.3 & 017.499632 & $-$01.032308 & 2001.8897 & 21.985$\pm$0.066 & 21.724$\pm$0.118 & $+$215.4$\pm$23.9 & $-$182.3$\pm$20.3 & 22 \\ 
011014.30$+$010619.1 & 017.559587 & $+$01.105316 & 2001.8897 & 21.920$\pm$0.044 & 19.692$\pm$0.024 & $+$546.2$\pm$12.8 & $+$41.4$\pm$15.5 & 24 \\ 
011755.09$+$005220.0$^{\star}$ & 019.479545 & $+$00.872235 & 2003.8809 & 22.543$\pm$0.121 & 20.919$\pm$0.105 & ($+$173.4$\pm$33.6) & ($+$21.9$\pm$30.6) & 8$^{\dagger}$ \\ 
013415.85$+$001456.0 & 023.566075 & $+$00.248906 & 2001.8897 & 21.794$\pm$0.026 & 20.453$\pm$0.048 & $+$77.9$\pm$17.5 & $-$105.6$\pm$20.6 & 28 \\ 
014956.27$+$001647.8 & 027.484491 & $+$00.279958 & 2001.8898 & 21.412$\pm$0.031 & 19.617$\pm$0.019 & $+$176.2$\pm$09.0 & $+$9.4$\pm$07.2 & 30 \\ 
020508.04$+$002458.7 & 031.283514 & $+$00.416331 & 2002.6787 & 21.257$\pm$0.023 & 21.212$\pm$0.072 & $+$184.5$\pm$10.1 & $-$28.2$\pm$05.8 & 44 \\ 
021546.76$+$010019.2 & 033.944852 & $+$01.005350 & 2002.6787 & 21.906$\pm$0.029 & 21.242$\pm$0.108 & $-$145.0$\pm$27.3 & $-$91.9$\pm$27.6 & 15 \\ 
024958.88$+$010624.3 & 042.495349 & $+$01.106755 & 2001.8899 & 21.728$\pm$0.036 & 20.837$\pm$0.079 & $+$106.5$\pm$16.8 & $-$101.0$\pm$09.7 & 29 \\ 
025316.06$+$005157.1 & 043.316948 & $+$00.865881 & 2002.6788 & 22.036$\pm$0.050 & 21.220$\pm$0.088 & $+$158.0$\pm$37.6 & $-$24.1$\pm$19.6 & 11 \\ 
032129.20$+$003212.8 & 050.371703 & $+$00.536911 & 2002.6789 & 21.446$\pm$0.025 & 21.305$\pm$0.093 & $+$120.2$\pm$10.0 & $-$134.4$\pm$08.0 & 21 \\ 
033203.57$+$003658.0$^{\star}$ & 053.014867 & $+$00.616101 & 2003.7330 & 22.272$\pm$0.054 & 21.223$\pm$0.130 & ($+$164.1$\pm$26.3) & ($+$1.3$\pm$13.1) & 8$^{\dagger}$ \\ 
033456.32$+$010618.7 & 053.734696 & $+$01.105197 & 2001.8900 & 21.048$\pm$0.024 & 19.334$\pm$0.015 & $+$387.0$\pm$07.5 & $-$107.6$\pm$06.0 & 32 \\ 
204821.28$-$004734.1 & 312.088698 & $-$00.792822 & 2001.7200 & 21.558$\pm$0.057 & 20.734$\pm$0.166 & ($-$255.8$\pm$28.6) & ($+$311.8$\pm$30.3) & 7$^{\dagger}$ \\ 
213151.87$-$001432.4 & 322.966158 & $-$00.242351 & 2001.7201 & 21.521$\pm$0.027 & 19.742$\pm$0.032 & $+$17.0$\pm$21.3 & $-$150.9$\pm$24.3 & 12 \\ 
215322.18$-$004553.9 & 328.342427 & $-$00.764994 & 2002.7768 & 22.453$\pm$0.228 & 21.178$\pm$0.140 & ($+$43.1$\pm$32.6) & ($-$278.6$\pm$56.8) & 6$^{\dagger}$ \\ 
215351.20$+$010120.3 & 328.463332 & $+$01.022322 & 2002.7768 & 21.884$\pm$0.042 & 20.965$\pm$0.149 & ($+$53.0$\pm$15.5) & ($-$144.3$\pm$35.2) & 10$^{\dagger}$ \\ 
215515.49$+$005128.0 & 328.814572 & $+$00.857790 & 2002.7768 & 21.879$\pm$0.060 & 20.777$\pm$0.076 & ($+$133.4$\pm$22.3) & ($-$63.5$\pm$22.0) & 11$^{\dagger}$ \\ 
215817.69$+$000300.3 & 329.573715 & $+$00.050106 & 2002.7768 & 21.717$\pm$0.055 & 19.822$\pm$0.039 & $+$88.5$\pm$16.6 & $-$117.0$\pm$18.9 & 13 \\ 
215934.25$+$005308.5 & 329.892736 & $+$00.885709 & 2002.7768 & 22.232$\pm$0.101 & 20.396$\pm$0.052 & $+$191.8$\pm$28.9 & $-$50.4$\pm$20.6 & 11 \\ 
220952.49$+$003325.2$^{\star}$ & 332.468692 & $+$00.557009 & 2004.8391 & 21.928$\pm$0.055 & 21.055$\pm$0.199 & ($+$94.3$\pm$33.9) & ($-$118.0$\pm$25.6) & 6$^{\dagger}$ \\ 
221911.35$+$010220.5 & 334.797303 & $+$01.039031 & 2002.7769 & 21.667$\pm$0.032 & 21.341$\pm$0.140 & ($+$200.6$\pm$38.6) & ($-$32.1$\pm$81.7) & 9$^{\dagger}$ \\ 
222939.14$+$010405.5 & 337.413111 & $+$01.068200 & 2001.7885 & 21.547$\pm$0.054 & 21.888$\pm$0.170 & $+$177.5$\pm$11.8 & $-$25.7$\pm$14.7 & 15 \\ 
225903.29$-$004154.2 & 344.763708 & $-$00.698389 & 2002.7769 & 22.080$\pm$0.069 & 20.305$\pm$0.062 & ($-$51.8$\pm$44.0) & ($-$249.0$\pm$29.9) & 8$^{\dagger}$ \\ 
230722.58$-$005746.6 & 346.844120 & $-$00.962948 & 2001.7886 & 22.189$\pm$0.052 & 20.183$\pm$0.031 & $+$129.1$\pm$21.7 & $-$55.3$\pm$13.9 & 23 \\ 
231407.82$+$004908.2 & 348.532598 & $+$00.818961 & 2001.7886 & 21.736$\pm$0.055 & 20.120$\pm$0.029 & $+$103.8$\pm$14.8 & $-$140.7$\pm$13.2 & 25 \\ 
231914.40$+$005615.9 & 349.810022 & $+$00.937755 & 2002.6784 & 22.061$\pm$0.095 & 21.406$\pm$0.241 & ($+$289.2$\pm$98.9) & ($-$422.8$\pm$152.) & 5$^{\dagger}$ \\ 
232935.99$-$011215.3 & 352.399971 & $-$01.204256 & 2002.6784 & 22.073$\pm$0.074 & 20.270$\pm$0.034 & $+$143.0$\pm$25.8 & $-$27.7$\pm$19.1 & 14 \\ 
234841.38$-$004022.1 & 357.172419 & $-$00.672832 & 2002.6784 & 21.999$\pm$0.051 & 19.932$\pm$0.052 & $+$108.9$\pm$16.0 & $-$132.1$\pm$18.7 & 15 \\ 
235826.48$+$003226.9 & 359.610344 & $+$00.540813 & 2002.6785 & 22.218$\pm$0.068 & 20.448$\pm$0.044 & $+$179.9$\pm$31.7 & $-$14.7$\pm$21.3 & 11 \\ 
235835.45$-$000909.5 & 359.647731 & $-$00.152658 & 2001.7887 & 21.120$\pm$0.020 & 19.252$\pm$0.014 & $-$96.5$\pm$10.2 & $-$228.7$\pm$09.6 & 20 \\ 
235841.98$+$000622.0 & 359.674927 & $+$00.106135 & 2002.6785 & 21.975$\pm$0.049 & 20.129$\pm$0.046 & $+$168.9$\pm$27.5 & $-$80.9$\pm$20.3 & 11 \\ 
\hline
\end{tabular}

\smallskip

\scriptsize{
Notes: Names, coordinates and epochs are taken from SDSS data release 6 (DR6; Adelman-McCarthy et al.~\cite{adelman08}) except for three objects
marked by $^{\star}$, for which due to the lack of DR6 data the latest available (in our analysis) epoch position from S82 is given instead.
Objects marked by $^{\dagger}$ in column (9) had observations spread over only 2-4 different epochs or less than about 4 years,
thus their proper motions (putted in parentheses) are uncertain.
}
\end{table*}

%
\begin{table*}
\caption{Known L and T dwarfs in S82 and their new 
SDSS-based mean $i$ and $z$ magnitudes 
and proper motions}
\label{tabLTpm}
\centering
\begin{tabular}{c c c c c c r r r}
\hline\hline
Discovery name               &   $J$ & opt. &  IR  &  mean $i$ & mean $z$ &$\mu_{\alpha}\cos{\delta}$ & $\mu_{\delta}$ & $N_{ep}$ \\
($^{\#}=$ comparison objects) & [mag]& SpT  &  SpT &[mag]   & [mag]  & [mas/yr]       &  [mas/yr]      & \\
(1)                          & (2)   & (3)  & (4)  & (5)    & (6)    & (7)            & (8)            & (9) \\
\hline
  SDSS J001608.44$-$004302.3$^1$& 16.326$\pm$0.116 &      &   L5.5$^1$&21.276$\pm$0.024 & 19.213$\pm$0.029&  $+$129.2$\pm$08.8&  $-$16.7$\pm$10.2& 11 \\
  SDSS J001911.65$+$003017.8$^2$& 14.921$\pm$0.037 &   L1$^2$ &          &19.353$\pm$0.014 &17.534$\pm$0.006 &   $-$17.7$\pm$08.9&  $-$59.1$\pm$11.8& 12 \\
  ULAS J002422.94$+$002247.9$^3$& 18.160$\pm$0.070 &          &    T4.5$^3$&             &             &             &            & \\
  ULAS J003402.77$-$005206.7$^4$& 18.140$\pm$0.080 &          &    T8.5$^4$&             &             &             &            & \\
  2MASS J00521232$+$0012172$^{5,\#}$ & 16.361$\pm$0.108 & & L2p$\pm$1$^5$&21.651$\pm$0.025 & 19.401$\pm$0.016&  $-$161.0$\pm$10.0& $-$186.2$\pm$07.9& 32 \\ 
 SDSSp J005406.55$-$003101.8$^{6,\#}$ & 15.731$\pm$0.053 & L1$^2$ &       &20.131$\pm$0.011 & 18.214$\pm$0.006&  $+$235.5$\pm$11.2& $-$179.5$\pm$06.6& 21 \\ 
 CFBDS J005910.90$-$011401.3$^7$& 18.060$\pm$0.030 &   &   T9$^7$&             &             &             &            & \\
   2MASSI J0104075$-$005328$^{8,\#}$ & 16.531$\pm$0.130 & L4.5$^{15}$ & &21.578$\pm$0.038 & 19.334$\pm$0.024&  $+$453.1$\pm$08.9&  $-$10.1$\pm$10.1& 15 \\ 
 SDSSp J010752.33$+$004156.1$^{9,\#}$ & 15.824$\pm$0.058 & L8$^2$ & L5.5$^1$&21.331$\pm$0.028 & 18.650$\pm$0.010& $+$637.5$\pm$06.0& $+$81.0$\pm$05.7& 24 \\ 
  ULAS J013117.53$-$003128.6$^{10}$& 18.600$\pm$0.040 &   &    T3$^{10}$&             &            &             &            & \\
  ULAS J013939.77$+$004813.8$^{10}$& 18.430$\pm$0.040 &   &   T7.5$^{10}$&             &            &             &            & \\
  SDSS J020333.26$-$010812.5$^1$& 16.858 &      &   L9.5$^1$&23.889$\pm$0.324 & 20.450$\pm$0.043&  $+$361.0$\pm$18.3&  $+$5.2$\pm$18.8& 13 \\
  ULAS J020336.94$-$010231.1$^3$& 18.040$\pm$0.050 &      &    T5$^3$&             &            &             &            & \\
  SDSS J020742.48$+$000056.2$^9$& 16.799$\pm$0.156 &      &   T4.5$^{13}$&24.372$\pm$0.266 & 20.258$\pm$0.038&  $+$143.8$\pm$15.8&  $-$28.2$\pm$10.5& 20 \\
        IfA 0230$-$Z1$^{11}$      & 18.200 &    &    T3$^{13}$& 23.414:  & 21.104:        & ($+$430:)   & ($+$325:)           & 1$^{\dagger}$ \\
  SDSS J022723.77$-$005518.6$^2$& 16.877 &   L0$^2$ &         &21.013$\pm$0.017 & 19.341$\pm$0.012&   $+$46.4$\pm$06.8&   $-$9.5$\pm$07.1& 32 \\
 SDSSp J023617.93$+$004855.0$^{9,\#}$ & 16.098$\pm$0.077 &L6$^2$ & L6.5$^1$&21.501$\pm$0.029 & 18.816$\pm$0.008& $+$133.9$\pm$05.7& $-$153.7$\pm$06.5& 32 \\ 
  SDSS J025601.86$+$011047.2$^2$& 16.212$\pm$0.102 &  L0$^2$ &      &20.502$\pm$0.011 & 18.719$\pm$0.009&  $+$125.7$\pm$13.0&  $-$69.0$\pm$14.7& 33 \\
 SDSSp J030136.53$+$002057.9$^6$& 16.882$\pm$0.190 &  L1:$^6$ &     &21.012$\pm$0.014 & 19.199$\pm$0.014&   $+$73.4$\pm$06.3&  $-$77.9$\pm$07.2& 31 \\
 SDSSp J030321.24$-$000938.2$^6$& 16.120$\pm$0.072 &  L0$^{6,*}$ &      &20.184$\pm$0.010 & 18.308$\pm$0.006&   $+$12.4$\pm$11.2&   $+$9.7$\pm$08.6& 32 \\
 SDSSp J032817.38$+$003257.2$^6$& 15.988$\pm$0.092 &  L3$^2$ &      &20.587$\pm$0.010 & 18.798$\pm$0.011&  $+$171.1$\pm$06.5&  $+$17.5$\pm$07.4& 22 \\
 SDSSp J033017.77$+$000047.8$^6$& 16.520$\pm$0.111 &  L0:$^6$ &     &20.637$\pm$0.017 & 18.861$\pm$0.010&   $-$38.2$\pm$13.6&  $-$75.4$\pm$15.4& 21 \\
 SDSSp J033035.13$-$002534.5$^{12}$& 15.311$\pm$0.050 &   L4$^2$    &   &20.155$\pm$0.007 & 18.017$\pm$0.008&  $+$392.4$\pm$10.8& $-$352.4$\pm$07.2& 25 \\
  ULAS J034832.97$-$001258.3$^{10}$& 18.720$\pm$0.040 &  early T$^{10}$   &   &             &             &             &            & \\
  SDSS J035448.73$-$002742.1$^1$& 17.282$\pm$0.211 &  &  L2$\pm$1$^1$&21.513$\pm$0.017 & 19.554$\pm$0.016&   $+$45.9$\pm$08.0&  $-$42.9$\pm$05.1& 27 \\
  SDSS J202820.32$+$005226.5$^2$& 14.298$\pm$0.035 &  L3$^2$ &     &18.787$\pm$0.022 & 16.997$\pm$0.001&  $+$135.3$\pm$18.0&  $+$14.7$\pm$08.1&  4 \\
  SDSS J212413.89$+$010000.3$^1$& 16.031$\pm$0.074 &   &     T5$^{13}$&23.701$\pm$0.573 & 19.645$\pm$0.046&  $+$237.6$\pm$31.6& $+$243.5$\pm$17.5& 12 \\
  SDSS J214046.55$+$011259.7$^{2,\#}$ & 15.891$\pm$0.084 & L3$^2$   &   &21.063$\pm$0.032 & 18.965$\pm$0.017&   $-$71.6$\pm$11.1& $-$205.1$\pm$05.2& 15 \\ 
  ULAS J222958.30$+$010217.2$^{10}$& 17.880$\pm$0.020 &    &    T2.5$^{10}$&             &             &             &            & \\
  ULAS J223955.76$+$003252.6$^3$& 18.850$\pm$0.050 &       &    T5.5$^3$&             &             &             &            & \\
 SDSSp J224953.45$+$004404.2$^9$& 16.587$\pm$0.125 &  L3$^2$ &  L5$\pm$1.5$^1$&21.635$\pm$0.045 & 19.477$\pm$0.021&   $+$88.8$\pm$09.8&  $+$17.8$\pm$09.2& 16 \\
 SDSSp J225529.09$-$003433.4$^6$& 15.650$\pm$0.056 &  L0:$^6$ &     &19.814$\pm$0.008 & 17.925$\pm$0.007&   $-$48.7$\pm$14.5& $-$179.9$\pm$15.6& 17 \\
  SDSS J225913.88$-$005158.2$^2$& 16.357$\pm$0.098 &  L2$^2$ &      &20.757$\pm$0.016 & 18.985$\pm$0.013&   $+$46.1$\pm$09.8&  $+$29.2$\pm$12.7& 22 \\
  2MASS J23453903$+$0055137$^{14}$& 13.771$\pm$0.030 &  L0$^{16}$   &   &18.001$\pm$0.003 & 16.171$\pm$0.005&   $+$93.4$\pm$06.8&  $-$67.1$\pm$11.6& 16 \\
\hline
\end{tabular}

\smallskip

\scriptsize{
References:
$^{1}$ Knapp et al.~(\cite{knapp04});
$^{2}$ Hawley et al.~(\cite{hawley02});
$^{3}$ Lodieu et al.~(\cite{lodieu07});
$^{4}$ Warren et al.~(\cite{warren07});
$^{5}$ Metchev et al.~(\cite{metchev08});
$^{6}$ Schneider et al.~(\cite{schneider02});
$^{7}$ Delorme et al.~(\cite{delorme08});
$^{8}$ Berriman et al.~(\cite{berriman03});
$^{9}$ Geballe et al.~(\cite{geballe02});
$^{10}$ Chiu et al.~(\cite{chiu08});
$^{11}$ Liu et al.~(\cite{liu02});
$^{12}$ Fan et al.~(\cite{fan00});
$^{13}$ Burgasser et al.~(\cite{burgasser06a});
$^{14}$ Reid et al. (in prep);
$^{15}$ Kirkpatrick et al. (in prep);
$^{16}$ unpublished;
$^*$ may be a distant red giant according to our non-significant proper motion;
$^{\dagger}$ uncertain proper motion (see text)
}
\end{table*}

\subsection{reduced proper motion diagram and $i$-magnitude cut}
\label{rpmd}

A reduced proper motion diagram (Fig.~\ref{fig1}) is often
used to separate the bottom of the main sequence (MS), the cool subdwarf (sd)
sequence and the CWD sequence,
when trigonometric distances and absolute magnitudes are not yet available
(see e.g. Kilic et al.~\cite{kilic06};
Carollo et al.~\cite{carollo06}; Vidrih et al.~\cite{vidrih07}).
The reduced proper motion 
(Hertzsprung \cite{hertzsprung05}; Luyten~\cite{luyten22}) 
is defined as $H=m+5+5\log{\mu}$, where $m$
is the apparent magnitude and $\mu$ is the proper motion expressed
in arcsec/yr. $H$ can be used as a statistical equivalent of
the absolute magnitude of a star in a sample with similar kinematics.

In Fig.~\ref{fig1} we have chosen an $H_z / (i-z)$ diagram in order to properly
represent
the L and T dwarfs, which are often barely detected in bluer SDSS passbands.
The $>$1000 S82 HPM objects with $\mu$$>$0.14~arcsec/yr found in step 1) 
of our survey (based on a limited number of
epochs) with formal errors of less than 20~mas/yr are shown as dots.
Some of these objects do in fact have spurious proper motions
despite their formal small errors, as our later checking has shown. 
Three sequences (MS, sd, wd) are well separated. We were aiming at optical
classification spectroscopy of the faintest new objects in all three
sequences. For this reason, and in order to make good use of VLT 
observing time, we decided to use an $i$-magnitude cut for our
targets. The crosses represent the 38 faint targets (Tab.~\ref{tabnewhpm})
finally selected after steps 2) and 3) of our HPM survey.
These are all new HPM sources with $i$$>$21.5, the majority of which
have proper motions based on less than 15 individual epoch measurements, 
as well as some
brighter objects (21.0$<$$i$$<$21.5) with typically more than 20 epochs. 
Tab.~\ref{tabnewhpm} lists accurate positions of our targets (finding charts
can be obtained from the public SDSS DR6 
webpage\footnote{http://cas.sdss.org/astrodr6/en/tools/chart/chart.asp}),
mean $i$ and $z$ magnitudes along with their errors, and the proper motion
components with their errors together with the number of epochs used for
their determination. For one of our HPM discoveries (SDSS J013415.85$+$001456.0)
the final proper motion solution (after excluding one outlier) led to a
total value of about 0.13~arcsec/yr. 
All other new HPM objects have final proper 
motions larger than 0.14~arcsec/yr. 

In the last column of Tab.~\ref{tabnewhpm} we have marked those objects which have 
only few observations effectively distributed over only 2-4 different
epochs or spread over a total time baseline of less than about four years.
The very large proper motion errors
of SDSS J231914.40$+$005615.9 are due to only five positions measured in
only two different seasons 
(two in autumn 2002 plus three in autumn 2003). However,
both proper motion components are almost three times larger than the errors.
This object is that with the largest value of the reduced proper motion
shown in Fig.~\ref{fig1} ($H_z\approx25$).
All objects marked in the last column of Tab.~\ref{tabnewhpm} are 
overplotted by large open circles in Fig.~\ref{fig1}. They occupy in particular
the region of suspected subdwarfs. Although the marked objecs as well as some
others with proper motion errors larger than $\pm$25~mas/yr passed our 3-step 
procedure, we may consider their large proper motions as relatively uncertain.

Other symbols shown in Fig.~\ref{fig1} represent the coolest known (in January 2006)
objects of different classes with available SDSS data: L and T dwarfs
from Gelino et al.~(\cite{gelino08}; and references therein), which all 
have $i-z$ colour indices larger than 1.5 (dotted line),
M9 dwarfs from Hawley et al.~(\cite{hawley02}; their proper motions 
were determined by us),
three cool subdwarfs from L{\'e}pine et al.~(\cite{lepine03,lepine04})
and Sivarani et al. (\cite{sivarani04}; see also 
Burgasser et al.~\cite{burgasser07}),
and two samples of CWDs discovered in SDSS, i.e. 
four CWDs with $\mu$$>$0.14~arcsec/yr from Gates et al.~(\cite{gates04})
and 7 CWDs with $T_{eff}$$<$4000~K and $\mu$$>$0.14~arcsec/yr from
Kilic et al.~(\cite{kilic06}).

One of the suspected L subdwarfs (the unpublished Sivarani object listed 
as sdL4: by Burgasser et al.~\cite{burgasser07}) falls in the region
of the L/T dwarfs with $i-z>1.5$, the other one (LSR 1610$-$0040, sdL:; 
L{\'e}pine et al.~\cite{lepine03}), described by
Cushing \& Vacca~(\cite{cushing06}) as a very peculiar object (M6p/sdM),
shows a much bluer colour
typical of mid-M dwarfs. Very recently, a trigonometric parallax was measured
for LSR 1610$-$0040 and it was found to be an astrometric binary of the 
Galactic halo population consisting of a mildly metal-poor M dwarf and a 
possible
substellar companion (Dahn et al.~\cite{dahn08}). The esdM6.5 object 
LSR J0822$+$1700 (L{\'e}pine et al.~\cite{lepine04}) is much bluer and 
lies right on the subdwarf sequence.

Half of our new targets are L dwarf candidates located in the region
right of the dotted line ($i-z>1.5$) in Fig.~\ref{fig1}. From their position 
in the reduced proper motion diagram, there 
could also be 
a T dwarf
among these candidates, since the known three objects exhibit a large spread
in the $i-z$ colour indices. Although M8 and M9 dwarfs 
also have
average $i-z$ colour indices between 1.6 and 1.7
(West et al.~\cite{west05}), we do not expect a
large contamination of our sample by late-M dwarfs, since the M9 dwarfs
with similarly large proper motions 
lie clearly above all our selected faint VLT targets in the reduced
proper motion diagram.
Our candidates with $0.5 < i-z < 1.5$ show a wide-spread
apparent extension of the subdwarf sequence.
On the blue side ($i-z<0.5$) there are some good candidates
for new benchmark sources among the coolest white dwarfs.

\subsection{comparison objects}
\label{compobj}

In Tab.~\ref{tabLTpm} we list all L and T dwarfs discovered until now
in the S82 region
from different surveys, not just the SDSS. 
The first column gives the discovery name marked
with the corresponding reference (comparison
objects selected by us as additional targets for our spectroscopic
observations are marked by $^{\#}$), the following columns list
$J$ magnitudes (2), optical spectral types with references (3),
near-infrared spectral types with references (4) (all data taken
from Gelino et al.~\cite{gelino08}, where the
$J$ magnitudes of all objects brighter than 17.5 are 
from 2MASS, whereas fainter $J$ magnitudes are from 
the discovery papers).
In the last five columns we 
give the mean $i$ and $z$ magnitudes (5,6) and the
proper motion components (7,8) derived by us using the number of 
SDSS epochs listed in the last column (9). 

At the beginning of our HPM survey (November 2005), there were
24 known L/T dwarfs falling in the S82 area.
Out of these we were able to re-discover all but two as HPM objects
(Tab.~\ref{tabLTpm}).
One of the latter (SDSSp J030321.24$-$000938.2) was classified by Schneider
et al.~(\cite{schneider02}) as an L0 dwarf, which has however 
according to our measurements a zero proper motion (we speculate
this may be in fact a distant red giant). For the other, 
a faint T dwarf (IfA 0230$-$Z1), 
there was only only one epoch measurement available in S82 (with
$i=23.414, z=21.104$). The comparison 
of the SDSS position of IfA 0230$-$Z1 with the position given in the
discovery paper by Liu et al.~(\cite{liu02}) hints at a large proper motion
of ($\mu_{\alpha}\cos{\delta},\mu_{\delta}$)$\approx$($+430,+325$)~mas/yr.
If its spectral classification and our SDSS identification are correct, 
then this is a T dwarf with a large reduced proper motion ($H_z$=24.8)
and a relatively blue colour ($i-z=+2.30$; see Fig.~\ref{fig1}).
Many of the known early-L dwarfs have according to our measurements
proper motions smaller than 
0.14~arcsec/yr and are therefore not shown in Fig.~\ref{fig1}. 

Note that the number of L and T dwarfs 
coincident with the S82 area, 
which was already
one of the best investigated sky areas with respect to these object
classes, has further increased
since the beginning of our survey. The current number is 34 according to
the status of 22 April 2008 in the DwarfArchives 
(Gelino et al.~\cite{gelino08}). This corresponds to an eight times higher
surface density of L and T dwarfs in S82 compared to the 
formal all-sky value simply computed from the 649 currently
known objects (Gelino et al.~\cite{gelino08}).
One new L dwarf was discovered in S82 
by Metchev et al.~(\cite{metchev08}). This object was independently found
by us in our HPM survey and originally selected as a promising new target
for our spectroscopic observations. We use it now as an additional known
comparison object, although it has only a near-infrared spectral type. 
Eight T dwarfs were discovered by 
Warren et al.~(\cite{warren07}), Lodieu et al.~(\cite{lodieu07}), and
Chiu et al.~(\cite{chiu08}) from UKIDSS data.
One of the
coolest known brown dwarfs, 
discovered from a deep imaging survey 
with the Canada-France-Hawaii Telescope 
(Delorme et al.~\cite{delorme08}), is also located in S82. 
All the new T dwarfs are too faint ($J>17.5$) so that we were not
able to detect them in our SDSS-based HPM survey.

   \begin{figure*}[!]
   \centering
   \includegraphics[angle=0,width=17.5cm]{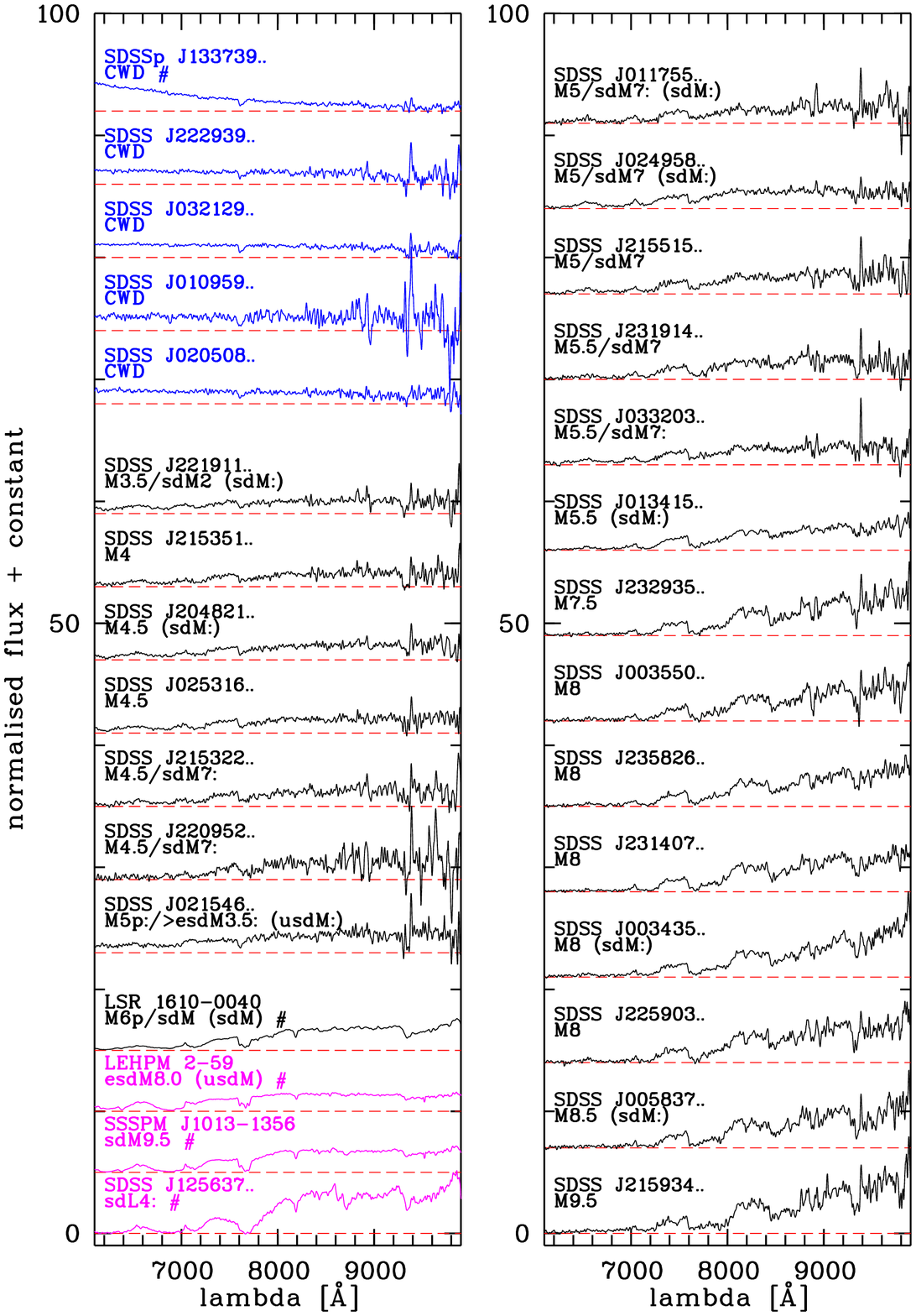}
   \caption{VLT/FORS1 spectra 
            of cool white dwarfs (five spectra in upper part of left column),
            comparison late-type (M/L) subdwarfs (four spectra in lower part of left
            column, including three ultracool $>$sdM7 subdwarfs), and
            M dwarfs/subdwarfs 
            sorted by spectral type (middle of left column
            and right column). Subdwarf candidates indicated in
            brackets were determined according to the revised metallicity classes
            (sdM, esdM, usdM) defined by L{\'e}pine et al.~(\cite{lepine07}).
            Comparison objects
            with previously known spectral types are marked by $\#$.
            All spectra are normalised at 7500\AA\, and smoothed with a
            running mean over 7 pixels.
               }
              \label{spec_cwdsdM}%
    \end{figure*}
%

   \begin{figure*}[!]
   \centering
   \includegraphics[angle=0,width=17.5cm]{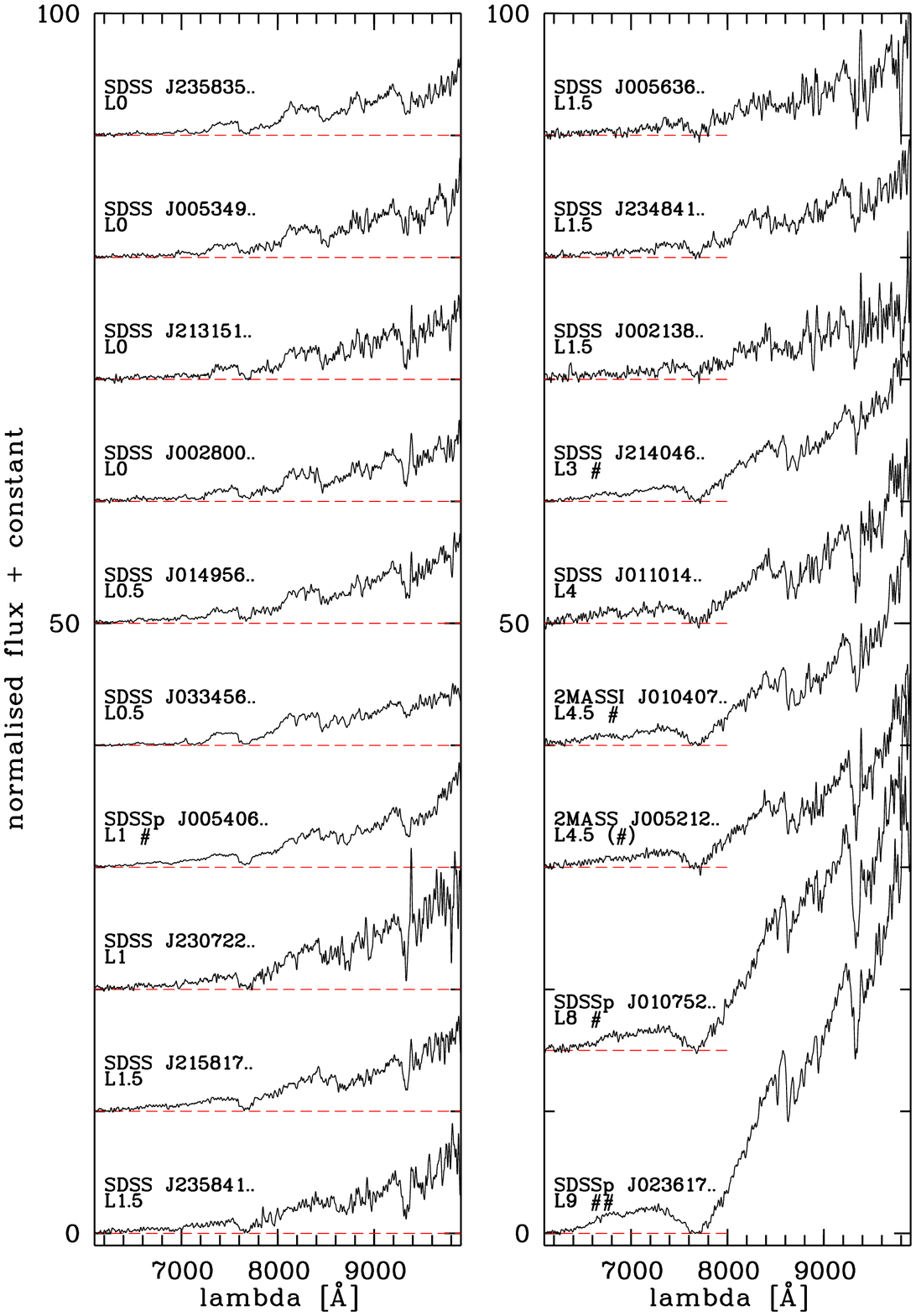}
   \caption{VLT/FORS1 spectra
            of L dwarfs
            sorted by spectral type (from top left to bottom right).
            Comparison objects
            with previously known spectral types are marked by $\#$.
            One comparison object with an available infrared but
            lacking optical spectral type is marked by ($\#$), The object
            marked by $\#\#$ has been re-classified by us as an L9
            dwarf. All spectra are normalised at 7500\AA\, and smoothed with a
            running mean over 7 pixels.
               }
              \label{spec_L}%
    \end{figure*}

In addition to the 6 known L dwarfs included as comparison objects
(Tab.~\ref{tabLTpm}), we selected four known cool subdwarfs as
''standard stars'' for our spectroscopic observations:
SDSS J125637.12$-$022452.4 (sdL4:; Sivarani et al.~\cite{sivarani04}; 
Burgasser et al.~\cite{burgasser07}),
LSR 1610$-$0040 
(M6p/sdM; 
L{\'e}pine et al.~\cite{lepine03};
Cushing \& Vacca~(\cite{cushing06}), 
SSSPM J1013$-$1356 (sdM9.5; Scholz et al.~\cite{scholz04a}), and
LEHPM 2-59 (esdM8; Pokorny et al.~\cite{pokorny04}; 
Burgasser \& Kirkpatrick \cite{burgasser06b}).
Finally, 
one of the coolest known white dwarfs was 
included in our spectroscopy target list:
SDSSp J133739.40+000142.8 (Harris et al.~\cite{harris01}).

%
\begin{table*}
\caption{Spectral indices measured for new HPM objects and comparison objects}
\label{tabspind}
\centering
\begin{tabular}{l r l r l r l r l r l r l r }
\hline\hline
Name                                      & PC3    & err   & TiO5 & err   & VOa  & err   & CaH1 & err   & CaH2 & err   & CaH3 & err   & $\zeta_{TiO/CaH}$ \\
(1)                                       & (2)    & (3)   & (4)  & (5)   & (6)  & (7)   & (8)  & (9)   & (10) & (11)  & (12) & (13)  & (14) \\
\hline
SDSS J002138.86$+$002605.8   &   2.35 &  0.23 & 1.35 &  0.49 & 2.10 &  0.46 & 3.42 &  3.86 & 1.07 &  0.41 & 2.29 &  0.93 & -3.5 \\
SDSS J002800.37$+$010252.0   &   2.35 &  0.11 & 0.43 &  0.07 & 2.17 &  0.19 & 4.51 &  4.59 & 0.37 &  0.07 & 0.52 &  0.06 & 0.77 \\
SDSS J003435.32$+$004633.9   &   1.95 &  0.04 & 0.28 &  0.07 & 2.23 &  0.15 & 0.45 &  0.51 & 0.22 &  0.07 & 0.46 &  0.06 & 0.82 \\
SDSS J003550.48$+$004631.3   &   1.68 &  0.10 & 0.05 &  0.09 & 2.40 &  0.22 & 0.66 &  0.48 & 0.17 &  0.09 & 0.58 &  0.06 & 1.15 \\
SDSS J005349.11$-$011354.4   &   2.20 &  0.11 & 0.63 &  0.06 & 2.66 &  0.22 & 0.87 &  0.66 & 0.46 &  0.08 & 1.06 &  0.20 & 1.47 \\
SDSS J005636.22$-$011131.9   &   4.25 &  0.67 & 0.98 &  0.24 & 2.46 &  0.35 & 0.97 &  0.45 & 0.12 &  0.14 & 0.58 &  0.24 & 0.03 \\
SDSS J005837.48$+$004435.6   &   1.72 &  0.09 & 0.39 &  0.04 & 2.48 &  0.20 & -74. & 799.5 & 0.22 &  0.06 & 0.44 &  0.07 & 0.69 \\
SDSS J010959.91$-$010156.3   &   1.03 &  0.11 & 1.08 &  0.11 & 2.17 &  0.22 & 1.11 &  0.22 & 1.03 &  0.12 & 1.12 &  0.12 & 0.48 \\
SDSS J011014.30$+$010619.1   &   6.03 &  0.85 & 2.12 &  0.47 & 1.75 &  0.16 & 0.70 &  0.43 & 1.21 &  0.25 & 0.95 &  0.18 & 6.65 \\
SDSS J011755.09$+$005220.0   &   1.17 &  0.05 & 0.55 &  0.03 & 2.24 &  0.15 & 1.05 &  0.30 & 0.26 &  0.06 & 0.62 &  0.05 & 0.60 \\
SDSS J013415.85$+$001456.0   &   1.42 &  0.03 & 0.43 &  0.06 & 2.18 &  0.10 & 1.09 &  0.36 & 0.30 &  0.05 & 0.52 &  0.03 & 0.73 \\
SDSS J014956.27$+$001647.8   &   2.45 &  0.11 & 0.55 &  0.14 & 2.48 &  0.15 & 1.27 &  0.18 & 0.56 &  0.10 & 0.69 &  0.12 & 0.94 \\
SDSS J020508.04$+$002458.7   &   0.91 &  0.07 & 0.96 &  0.06 & 2.18 &  0.12 & 0.99 &  0.08 & 1.16 &  0.06 & 1.09 &  0.06 & -0.2 \\
SDSS J021546.76$+$010019.2   &   1.35 &  0.07 & 1.06 &  0.03 & 2.02 &  0.14 & 0.76 &  0.09 & 0.60 &  0.05 & 0.62 &  0.03 & -0.1 \\
SDSS J024958.88$+$010624.3   &   1.29 &  0.05 & 0.32 &  0.02 & 1.92 &  0.11 & 0.54 &  0.15 & 0.24 &  0.02 & 0.40 &  0.03 & 0.75 \\
SDSS J025316.06$+$005157.1   &   1.01 &  0.04 & 0.36 &  0.02 & 2.12 &  0.13 & 0.78 &  0.04 & 0.40 &  0.03 & 0.81 &  0.04 & 1.27 \\
SDSS J032129.20$+$003212.8   &   1.06 &  0.04 & 1.03 &  0.03 & 1.86 &  0.10 & 0.96 &  0.02 & 1.02 &  0.02 & 0.93 &  0.03 & 0.55 \\
SDSS J033203.57$+$003658.0   &   1.15 &  0.05 & 0.45 &  0.04 & 2.33 &  0.15 & 0.73 &  0.13 & 0.33 &  0.03 & 0.84 &  0.05 & 1.03 \\
SDSS J033456.32$+$010618.7   &   2.56 &  0.06 & 0.10 &  0.03 & 2.00 &  0.09 & 0.75 &  0.60 & 0.21 &  0.03 & 0.25 &  0.03 & 0.91 \\
SDSS J204821.28$-$004734.1   &   1.14 &  0.05 & 0.62 &  0.04 & 2.17 &  0.07 & 1.40 &  0.14 & 0.53 &  0.04 & 0.66 &  0.04 & 0.72 \\
SDSS J213151.87$-$001432.4   &   2.14 &  0.12 & 0.97 &  0.36 & 2.64 &  0.28 & -3.5 & -4.39 & 0.39 &  0.21 & 0.81 &  0.26 & 0.06 \\
SDSS J215322.18$-$004553.9   &   1.09 &  0.07 & 0.31 &  0.04 & 1.93 &  0.12 & 1.21 &  0.46 & 0.48 &  0.03 & 0.68 &  0.06 & 1.27 \\
SDSS J215351.20$+$010120.3   &   1.02 &  0.04 & 0.38 &  0.04 & 1.89 &  0.12 & 0.91 &  0.26 & 0.37 &  0.07 & 0.55 &  0.09 & 0.86 \\
SDSS J215515.49$+$005128.0   &   1.32 &  0.06 & 0.24 &  0.02 & 1.86 &  0.13 & 0.73 &  0.19 & 0.33 &  0.05 & 0.53 &  0.04 & 1.00 \\
SDSS J215817.69$+$000300.3   &   2.78 &  0.13 & 0.89 &  0.08 & 2.11 &  0.10 & 0.51 &  0.37 & 0.74 &  0.07 & 0.71 &  0.07 & 0.34 \\
SDSS J215934.25$+$005308.5   &   1.92 &  0.10 & 0.36 &  0.09 & 2.49 &  0.19 & 1.37 &  0.32 & 0.36 &  0.08 & 0.57 &  0.12 & 0.90 \\
SDSS J220952.49$+$003325.2   &   1.10 &  0.12 & 0.46 &  0.10 & 1.99 &  0.30 & 1.21 &  0.50 & 0.69 &  0.10 & 0.53 &  0.11 & 1.08 \\
SDSS J221911.35$+$010220.5   &   1.16 &  0.06 & 0.74 &  0.09 & 2.09 &  0.12 & 0.69 &  0.09 & 0.52 &  0.05 & 0.85 &  0.04 & 0.68 \\
SDSS J222939.14$+$010405.5   &   0.93 &  0.04 & 1.04 &  0.06 & 1.80 &  0.10 & 0.83 &  0.08 & 1.04 &  0.05 & 1.00 &  0.07 & 0.35 \\
SDSS J225903.29$-$004154.2   &   2.06 &  0.09 & 0.12 &  0.03 & 2.43 &  0.12 & 1.08 &  0.42 & 0.09 &  0.06 & 0.36 &  0.08 & 0.89 \\
SDSS J230722.58$-$005746.6   &   2.64 &  0.15 & 0.59 &  0.09 & 2.44 &  0.40 & 2.66 &  1.77 & 0.39 &  0.07 & 0.11 &  0.04 & 0.43 \\
SDSS J231407.82$+$004908.2   &   1.95 &  0.09 & 0.26 &  0.07 & 2.34 &  0.11 & 1.10 &  0.29 & 0.31 &  0.06 & 0.40 &  0.09 & 0.86 \\
SDSS J231914.40$+$005615.9   &   1.34 &  0.09 & 0.37 &  0.07 & 1.97 &  0.14 & 1.18 &  0.33 & 0.40 &  0.06 & 0.57 &  0.05 & 0.92 \\
SDSS J232935.99$-$011215.3   &   1.99 &  0.09 & 0.06 &  0.05 & 2.10 &  0.13 & -0.5 &  0.35 & 0.25 &  0.07 & 0.55 &  0.12 & 1.17 \\
SDSS J234841.38$-$004022.1   &   3.03 &  0.19 & 1.00 &  0.10 & 2.13 &  0.19 & 0.22 &  0.24 & 0.78 &  0.12 & 0.69 &  0.07 & 0.00 \\
SDSS J235826.48$+$003226.9   &   1.72 &  0.08 & 0.33 &  0.06 & 2.34 &  0.13 & 0.82 &  0.35 & 0.16 &  0.05 & 0.67 &  0.10 & 0.85 \\
SDSS J235835.45$-$000909.5   &   2.17 &  0.10 & 0.17 &  0.03 & 2.28 &  0.12 & 1.35 &  0.53 & 0.44 &  0.03 & 0.71 &  0.04 & 1.51 \\
SDSS J235841.98$+$000622.0   &   2.82 &  0.23 & 0.75 &  0.17 & 1.96 &  0.26 & 1.00 &  0.27 & 0.65 &  0.09 & 0.74 &  0.05 & 0.70 \\
comparison L dwarfs in S82:\\
2MASS J00521232$+$0012172    &   4.79 &  0.55 & 1.86 &  0.21 & 1.78 &  0.15 & 1.33 &  0.33 & 1.25 &  0.22 & 1.19 &  0.20 & 3.22 \\
SDSSp J005406.55$-$003101.8  &   2.66 &  0.07 & 0.96 &  0.04 & 2.15 &  0.09 & 0.72 &  0.08 & 0.54 &  0.04 & 0.68 &  0.06 & 0.09 \\
2MASSI J0104075$-$005328     &   5.65 &  0.44 & 0.85 &  0.07 & 1.66 &  0.11 & 0.99 &  0.20 & 0.61 &  0.05 & 0.53 &  0.04 & 0.27 \\
SDSSp J010752.33$+$004156.1  &   7.30 &  0.72 & 1.00 &  0.05 & 2.09 &  0.19 & 2.29 &  2.48 & 1.03 &  0.08 & 0.92 &  0.06 & 0.03 \\
SDSSp J023617.93$+$004855.0  &  10.41 &  0.61 & 1.21 &  0.08 & 2.14 &  0.18 & 0.71 &  0.22 & 0.94 &  0.05 & 0.96 &  0.05 & 10.4 \\
SDSS J214046.55$+$011259.7   &   4.02 &  0.31 & 1.11 &  0.06 & 1.85 &  0.14 & 0.84 &  0.25 & 0.84 &  0.06 & 0.70 &  0.05 & -0.4 \\
comparison cool subdwarfs:\\
SDSS J125637.12$-$022452.4     &   3.60 &  0.12 & 0.27 &  0.01 & 1.82 &  0.07 & 0.32 &  0.05 & 0.15 &  0.01 & 0.19 &  0.02 & 0.71 \\
LSR 1610$-$0040              &   1.60 &  0.02 & 0.37 &  0.02 & 2.09 &  0.01 & 0.75 &  0.04 & 0.29 &  0.01 & 0.52 &  0.02 & 0.80 \\
SSSPM J1013$-$1356           &   1.69 &  0.02 & 0.36 &  0.02 & 2.08 &  0.02 & 0.37 &  0.03 & 0.12 &  0.01 & 0.22 &  0.01 & 0.63 \\
LEHPM 2-59                   &   1.34 &  0.02 & 0.84 &  0.05 & 2.11 &  0.03 & 0.37 &  0.02 & 0.20 &  0.01 & 0.29 &  0.01 & 0.17 \\
comparison CWD:\\
SDSSp J133739.40$+$000142.8  &   0.71 &  0.04 & 0.93 &  0.03 & 1.92 &  0.08 & 1.00 &  0.03 & 1.16 &  0.03 & 1.04 &  0.02 & -0.3 \\
\hline
\end{tabular}
\end{table*}

%
\begin{table*}
\caption{Spectral types, distances and tangential velocities of new HPM objects and comparison objects}
\label{tabsptdv}
\centering
\begin{tabular}{l r l r l r l r r c c c c}
\hline\hline
Name                                    & SpT   & err & SpT   & err & SpT   & err & SpT & SpT   & SpT    & sd ? & $d_{spec}$ & $v_{tan}$ \\
abbreviated & PC3 &     & TiO5  &     & VOa   &     & ind & comp  & adopted& \scriptsize{$\zeta_{TiO/CaH}$}& [pc]       & [km/s] \\
(1)                                     & (2)   & (3) & (4)  & (5)   & (6)  & (7) & (8) & (9)  & (10)     & (11)  & (12)       & (13)  \\
\hline
SDSS J002138..  &  M9.7 & 0.5 &  L3.9 & 2.8 &  L2.0 & 3.3 &  L1.9 & L1.5 &  L1.5  &      &   112 &   97 \\
SDSS J002800..  &  M9.7 & 0.2 &  M8.7 & 0.4 &  L1.4 & 1.3 &  M9.9 & L0.5 &  L0.0  &      &   151 &   119 \\
SDSS J003435..  &  M8.4 & 0.2 &  M7.8 & 0.4 &  M7.1 & 1.6 &  M7.8 & M8.5 &  M8.0  & sdM: &   197 &   134 \\
SDSS J003550..  &  M7.3 & 0.5 &  M7.7 & 1.0 &  M9.0 & 2.3 &  M8.0 & M7.5 &  M8.0  &      &   266 &   202 \\
SDSS J005349..  &  M9.3 & 0.3 &  M9.8 & 0.4 &  M8.0 & 1.6 &  M9.9 & L0.5 &  L0.0  &      &   149 &   102 \\
SDSS J005636..  &  L2.7 & 0.5 &  L1.8 & 1.4 &  M9.4 & 2.5 &  L1.3 & L1.5 &  L1.5  &      &   119 &    94 \\
SDSS J005837..  &  M7.5 & 0.4 &  M8.5 & 0.2 &  M9.3 & 1.4 &  M8.4 & M9.0 &  M8.5  & sdM: &   182 &   290 \\
SDSS J010959..  &  M3.2 & 0.8 & $<$M0 & 1.1 &  M6.5 & 2.3 &       & CWD  &  CWD   &      &   165$^{\star}$ & 221 \\
SDSS J011014..  &  L4.0 & 0.5 &  L8.3 & 2.7 &  L4.4 & 1.1 &  L4.2 & L3.5 &  L4.0  &      &    64 &   166 \\
SDSS J011755..  &  M4.3 & 0.3 &  M2.2 & 0.3 &  M7.3 & 1.6 &  M4.6 & M5.0 &  M5.0/sdM7:  & sdM: &   920/625 &   (762/518)$^{\dagger}$ \\
SDSS J013415..  &  M5.9 & 0.2 &  M3.6 & 0.6 &  M6.7 & 1.0 &  M5.4 & M6.0 &  M5.5  & sdM: &   612 &   381 \\
SDSS J014956..  &  M9.9 & 0.2 &  M9.4 & 0.8 &  M9.3 & 1.1 &  M9.5 & L1.0 &  L0.5  &      &   118 &    99 \\
SDSS J020508..  &  M2.3 & 0.5 & $<$M0 & 0.6 &  M6.6 & 1.3 &       & CWD  &  CWD   &      &   118$^{\star}$ & 104 \\
SDSS J021546..  &  M5.4 & 0.5 & $<$M0 & 0.3 &  M5.0 & 1.5 &       & esdM: &  M5p:/$>$esdM3.5:  & usdM:&  1068/$<$1138 &   869/$<$926 \\
SDSS J024958..  &  M5.0 & 0.3 &  M4.7 & 0.2 &  M3.9 & 1.2 &  M4.5 & M5.5 &  M5.0/sdM7  & sdM: &   886/602 &   617/419 \\
SDSS J025316..  &  M3.1 & 0.3 &  M4.3 & 0.3 &  M6.0 & 1.3 &  M4.5 & M4.5 &  M4.5  &      &  1387 &  1050 \\
SDSS J032129..  &  M3.4 & 0.3 & $<$M0 & 0.3 &  M3.2 & 1.0 &       & CWD  &  CWD   &      &   129$^{\star}$ & 110 \\
SDSS J033203..  &  M4.1 & 0.4 &  M3.4 & 0.5 &  M8.3 & 1.6 &  M5.2 & M5.5 &  M5.5/sdM7:  &      &   872/719 &   (678/559)$^{\dagger}$ \\
SDSS J033456..  &  L1.3 & 0.1 &  M6.8 & 0.2 &  L2.7 & 0.6 &  L0.3 & L0.5 &  L0.5  &      &   103 &   197 \\
SDSS J204821..  &  M4.0 & 0.4 &  M1.5 & 0.4 &  M6.5 & 0.7 &  M4.0 & M4.5 &  M4.5  & sdM: &  1109 &  (2119)$^{\dagger}$ \\
SDSS J213151..  &  M9.1 & 0.4 &  L1.7 & 2.1 &  M8.2 & 2.0 &  M9.7 & L0.5 &  L0.0  &      &   141 &   101 \\
SDSS J215322..  &  M3.7 & 0.5 &  M4.9 & 0.5 &  M4.0 & 1.3 &  M4.2 & M5.0 &  M4.5/sdM7:  &      &  1360/704 &  (1817/941)$^{\dagger}$ \\
SDSS J215351..  &  M3.2 & 0.3 &  M4.1 & 0.4 &  M3.6 & 1.2 &  M4.0 & M4.0 &  M4.0  &      &  1626 &  (1185)$^{\dagger}$ \\
SDSS J215515..  &  M5.2 & 0.4 &  M5.6 & 0.2 &  M3.2 & 1.3 &  M4.7 & M5.5 &  M5.0/sdM7  &      &   862/585 &   (604/410)$^{\dagger}$ \\
SDSS J215817..  &  L1.5 & 0.1 &  L1.3 & 0.4 &  L1.9 & 0.7 &  L1.6 & L1.0 &  L1.5  &      &   106 &    74 \\
SDSS J215934..  &  M8.3 & 0.4 &  M8.3 & 0.5 &  M9.2 & 1.4 &  M8.6 & L0.5 &  M9.5  &      &   206 &   193 \\
SDSS J220952..  &  M3.7 & 0.8 &  M3.3 & 1.0 &  M4.7 & 3.1 &  M3.9 & M5.5 &  M4.5/sdM7:  &      &  1285/665 &   (920/476)$^{\dagger}$ \\
SDSS J221911..  &  M4.2 & 0.4 &  M0.2 & 1.0 &  M5.7 & 1.3 &  M3.4 & M4.0 &  M3.5/sdM2:  & sdM: &  2548/1910 &  (2454/1839)$^{\dagger}$ \\
SDSS J222939..  &  M2.5 & 0.3 & $<$M0 & 0.7 &  M2.7 & 1.1 &       & CWD  &  CWD   &      &   135$^{\star}$ & 115 \\
SDSS J225903..  &  M8.9 & 0.3 &  M6.9 & 0.2 &  M9.3 & 1.2 &  M8.4 & M8.0 &  M8.0  &      &   265 &   (319)$^{\dagger}$ \\
SDSS J230722..  &  L1.3 & 0.1 &  M9.6 & 0.5 &  M9.6 & 2.8 &  L0.2 & L2.0 &  L1.0  &      &   136 &    91 \\
SDSS J231407..  &  M8.5 & 0.3 &  M7.7 & 0.4 &  M8.3 & 1.2 &  M8.2 & M8.0 &  M8.0  &      &   243 &   202 \\
SDSS J231914..  &  M5.4 & 0.5 &  M4.2 & 0.8 &  M4.4 & 1.5 &  M4.7 & M6.0 &  M5.5/sdM7  &      &   949/782 &  (2304/1899)$^{\dagger}$ \\
SDSS J232935..  &  M8.6 & 0.3 &  M6.6 & 0.3 &  M5.8 & 1.4 &  M7.0 & M8.0 &  M7.5  &      &   281 &   194 \\
SDSS J234841..  &  L1.7 & 0.2 &  L1.9 & 0.6 &  L1.7 & 1.3 &  L1.8 & L1.5 &  L1.5  &      &   112 &    91 \\
SDSS J235826..  &  M7.5 & 0.4 &  M8.1 & 0.3 &  M8.4 & 1.3 &  M8.0 & M7.5 &  M8.0  &      &   283 &   242 \\
SDSS J235835..  &  M9.2 & 0.3 &  M7.2 & 0.2 &  L0.7 & 0.9 &  M9.0 & L0.5 &  L0.0  &      &   112 &   132 \\
SDSS J235841..  &  L1.5 & 0.2 &  L0.5 & 1.0 &  L2.9 & 1.8 &  L1.6 & L1.0 &  L1.5  &      &   122 &   109 \\
comparison L\\
2MASS J005212..   &  L3.1 & 0.4 &  L6.8 & 1.2 &  L4.2 & 1.1 &  L4.7 & L4.5 &  L4.5  &      &    53 &    62 \\
SDSSp J005406.. &  L1.4 & 0.1 &  L1.6 & 0.2 &  L1.6 & 0.6 &  L1.5 &      &  L1.0  &      &    55 &    77 \\
2MASSI J010407..  &  L3.7 & 0.3 &  L1.0 & 0.4 &  L5.1 & 0.8 &  L3.3 &      &  L4.5  &      &    51 &   110 \\
SDSSp J010752.. &  L4.7 & 0.4 &  L1.9 & 0.3 &  L2.0 & 1.3 &       &      &  L8.0  &      &    17 &    51 \\
SDSSp J023617.. &  L5.8 & 0.1 &  L3.1 & 0.4 &  L1.7 & 1.3 &       & L9.0 &  L9.0  &      &    18 &    17 \\
SDSS J214046..  &  L2.5 & 0.2 &  L2.5 & 0.3 &  L3.7 & 1.0 &  L2.9 &      &  L3.0  &      &    54 &    56 \\
comparison sd\\
SDSS J125637..  &  L2.2 & 0.1 &  M7.8 & 0.1 &  L3.9 & 0.5 &       &       & sdL4:  & sdL  &   120$^1$ &   353$^1$ \\
LSR 1610$-$0040             &  M6.9 & 0.1 &  M4.2 & 0.2 &  M5.7 & 0.2 &       &       &  M6p/sdM   & sdM &   32$^2$ &   221$^2$  \\
SSSPM J1013..          &  M7.3 & 0.1 &  M8.3 & 0.1 &  L2.1 & 0.1 &       &       & sdM9.5 & sdM  &    50$^3$ &   244$^3$ \\
LEHPM 2-59                  &  M5.4 & 0.1 &  L1.0 & 0.3 &  M5.9 & 0.3 &       &       & esdM8  & usdM &    66$^4$ &   233$^4$ \\
comparison CWD\\
SDSSp J133739.. &  M0.6 & 0.3 & $<$M0 & 0.4 &  M3.9 & 0.8 &       &       &  CWD   &      &    20-54$^5$ &  17-46$^5$ \\
\hline
\end{tabular}

\smallskip

\scriptsize{
Notes: 
$^1$ - Sivarani et al.~(\cite{sivarani04}); Burgasser et al~(\cite{burgasser07}), 
$^2$ - according to Dahn et al~(\cite{dahn08}), 
$^3$ - Scholz et al.~(\cite{scholz04a}), 
$^4$ - Pokorny et al.~(\cite{pokorny04}), Burgasser \& Kirkpatrick (\cite{burgasser06b}), 
$^5$ - Harris et al.~(\cite{harris01}).
Distances in column (12) marked by $^{\star}$ were derived assuming an absolute magnitude of $M_i=15.9$;
uncertainties in distances and tangential velocities are 
$\sim$23\% for M7-L9 dwarfs and $\sim$46\% for new CWDs and M3-M6 dwarfs
(including alternative subdwarf types)
(see text). 
Objects with tangential velocities in column (13) marked by $^{\dagger}$ 
have uncertain proper motions (see Tab.~\ref{tabnewhpm}).
Their tangential velocities (putted in parentheses) have therefore an 
additional uncertainty, i.e. they may be overestimated by an unknown factor.
}
\end{table*}

   \begin{figure*}[!]
   \centering
   \includegraphics[angle=0,width=13.0cm]{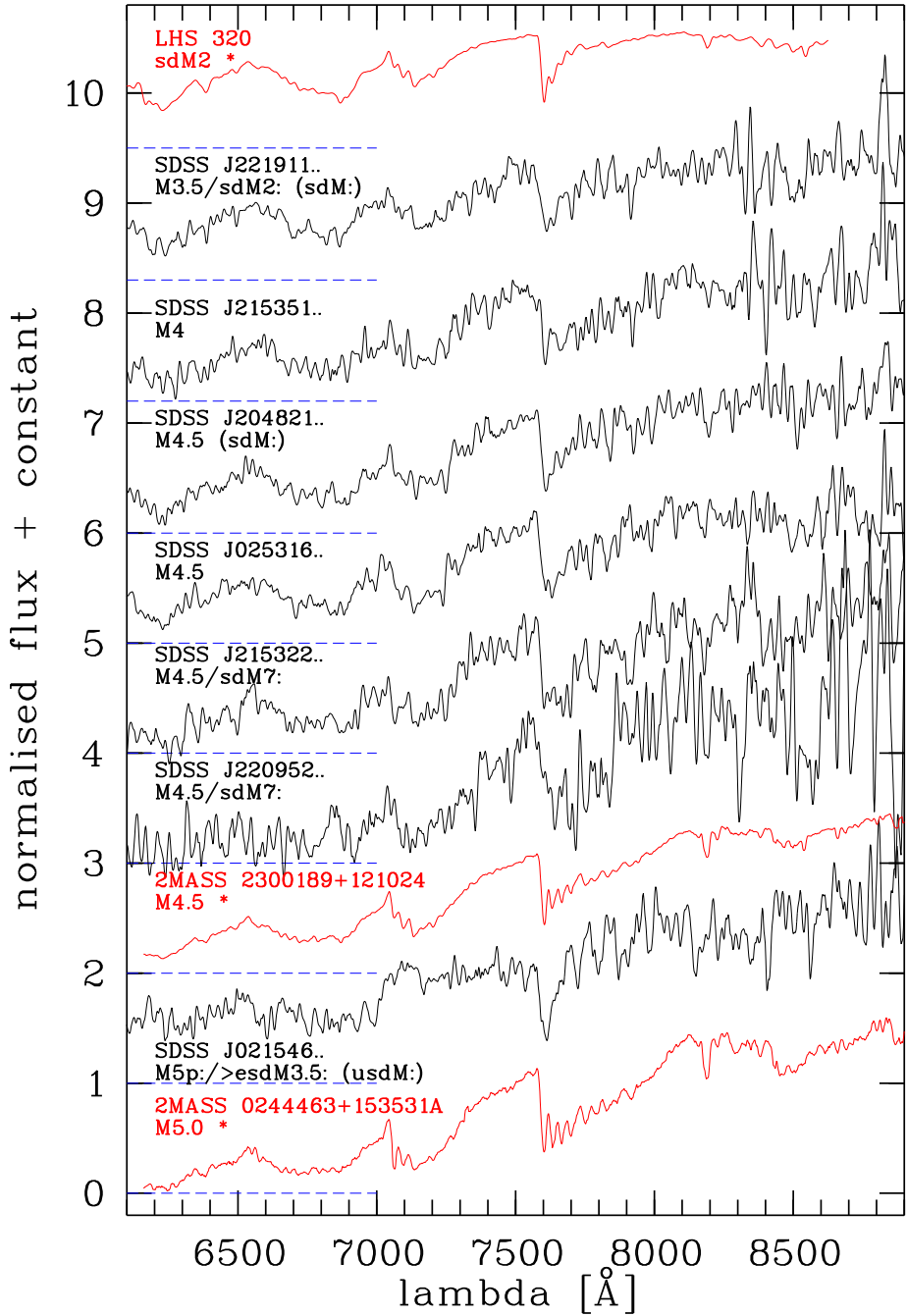}
   \caption{VLT/FORS1 spectra of all objects initially classified as 
            M3.5-M5p in comparison to Keck/LRIS spectra (marked by $*$)
            of known dwarfs and subdwarfs taken from Neill Reid's 
            webpage (Gizis~\cite{gizis97};
            Kirkpatrick et al.~\cite{kirkpatrick99}). 
            All spectra
            have been smoothed with a running mean over 7 pixels.
               }
              \label{plot_MsdM_1}%
    \end{figure*}

   \begin{figure*}[!]
   \centering
   \includegraphics[angle=0,width=13.0cm]{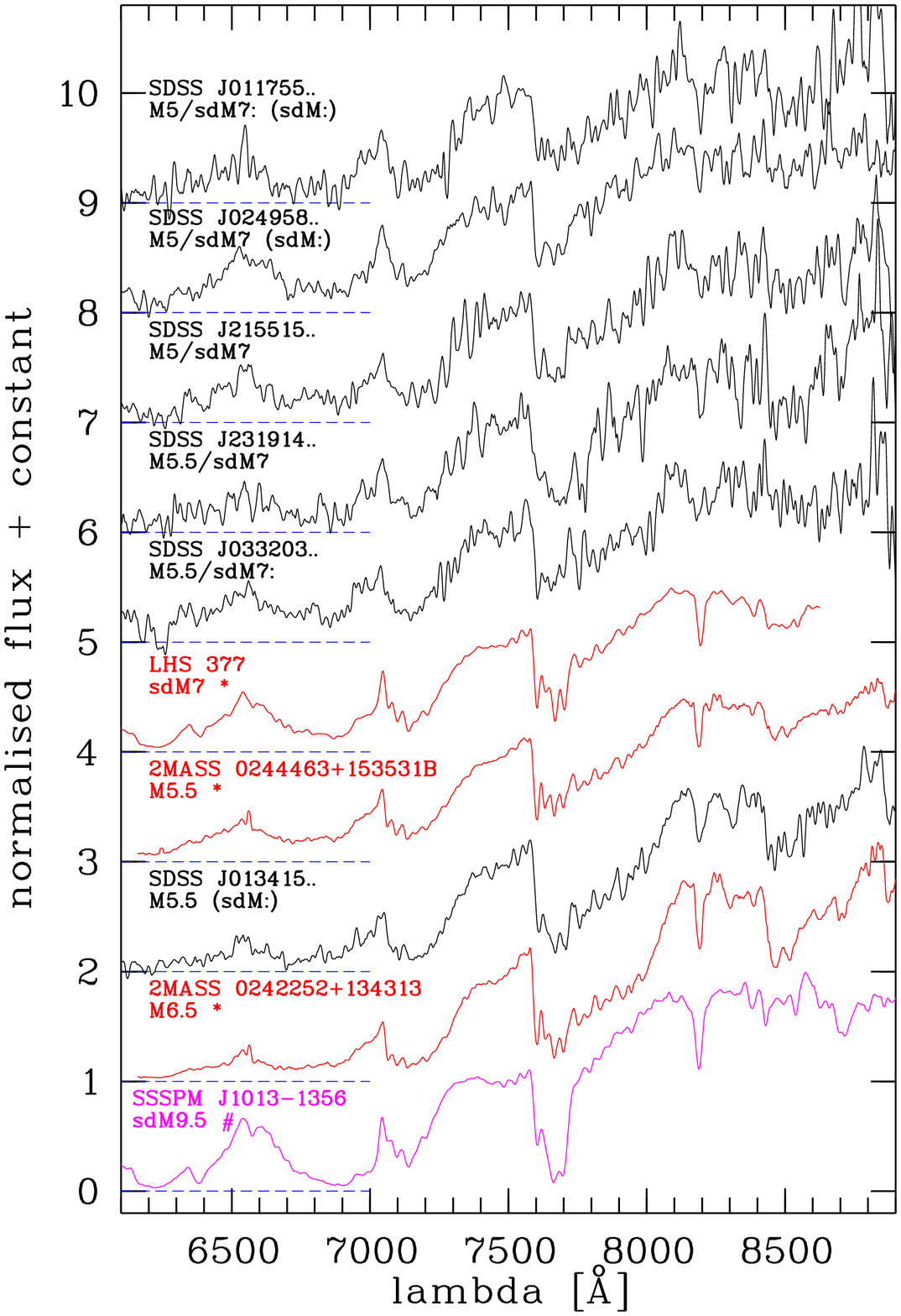}
   \caption{Same as Fig.~\ref{plot_MsdM_1} for all
           objects initially classified as M5-5.5. Also shown for comparison
           is the VLT/FORS1 spectrum 
           of SSSPM J1013$-$1356.
               }
              \label{plot_MsdM_2}%
    \end{figure*}

%

\section{Optical classification spectroscopy}
\label{sec3}

The existing spectroscopic classification schemes for L dwarfs, cool
subdwarfs and CWDs work best in the optical wavelength region. Therefore,
a first look in the optical seems to be justified in order to classify
our new, extremely faint HPM
discoveries. In the following
we describe the spectroscopic observations,
data reduction, and the classification of our targets based on spectral
indices and comparison with spectra of known objects of different classes.

\subsection{observations and data reduction}
\label{obsdata}

Low-resolution spectroscopy was obtained during 25 hours of service
observations
with the ESO VLT Keuyen telescope and the FORS1 instrument
(Appenzeller et al.~\cite{appenzeller98}) during period
78 (October 2006 to March 2007). We used grism GRIS\_150I+17 with
a wavelength range of 600-1100~nm, and a dispersion of
23~nm/mm. The slit, set to the parallactic angle, had a width of 1~arcsec,
and the seeing varied
between 0.45 and 1.4 arcsec with a couple of spectra with worse seeing.
With 44 faint targets, including 38
new objects (Tab.~\ref{tabnewhpm}) and 6 known L dwarfs (Tab.~\ref{tabLTpm})
the effective exposure times for the spectra ranged typically between
13 and 23 minutes. 
For 5 brighter
(16$<$$i$$<$19.5) comparison objects,
including 
one CWD 
and four late-type subdwarfs, only relatively
short exposure times of one to five minutes were needed.

The spectra were bias subtracted, flat-fielded and wavelength
calibrated before they were extracted using the optimal extraction
algorithm within the
IRAF\footnote{
IRAF is the Image Reduction and Analysis
Facility, made available to the astronomical community by the National
Optical Astronomy Observatories, which are operated by AURA, Inc., under
cooperative agreement with the National Science Foundation.}
{\tt onedspec} package.
  The spectra have been flux-calibrated using spectra of four standard
stars (Feige 100, LTT 2415, LTT 1020, \& LTT 3218) with
spectro-photometry from Hamuy et al. (\cite{hamuy94}) and using the CTIO
extinction curve provided
within IRAF. As the
spectra have been obtained on many different nights we do not have
appropriate nightly calibration spectra and due to the fairly low signal
to noise it was not reasonable to attempt to scale and subtract a
standard telluric spectrum from the data.

The spectra of all targets and comparison objects are normalised at 7500\AA\,
and are shown 
with the same scale
in Fig.~\ref{spec_cwdsdM} (CWDs, M (sub)dwarfs and ultracool
subdwarfs) and Fig.~\ref{spec_L} (L dwarfs), respectively sorted by spectral
types.
The spectra of our comparison objects are marked by $\#$.

\subsection{L and M spectral types}
\label{sptML}

For the spectral typing of our targets we have first used the visual 
comparison of their spectra with those of the known comparison objects.
A surprisingly large number of our targets looked like the 
peculiar (sub)dwarf LSR 1610$-$0040 (M6p/sdM)
while their continuum was not that red. 
Since we had not included normal mid- to late-type M dwarfs in our observations,
we selected in addition 
to our own template VLT/FORS1 spectra
a sequence of M dwarf spectra 
(observed with Keck/LRIS) 
from the spectral library provided by 
Neill Reid\footnote{http://www.stsci.edu/\~{}inr/ultracool.html}
for our comparison.

Although the signal-to-noise and resolution of our spectra 
are rather low, we
have also used the measurements of spectral indices in order to
classify our objects. In particular, we have measured the
TiO5 index defined by Reid et al.~(\cite{reid95}), the
VOa index defined by Kirkpatrick et al.~(\cite{kirkpatrick99}), and the
PC3 index defined by Mart\'{\i}n et al.~(\cite{martin99}).
The spectral indices together with their errors determined from the 
standard deviations of flux values in the corresponding wavelength
intervals are listed in Tab.~\ref{tabspind}. We have then applied the
spectral type versus spectral index relation for TiO5 and VOa given by
Cruz \& Reid (\cite{cruz02}) and for PC3 given by Mart\'{\i}n
et al.~(\cite{martin99}). The results are shown in Tab.~\ref{tabsptdv},
where the average spectral types based on spectral indices,
the spectral types based on visual comparison, and the finally adopted 
spectral types are shown in columns (8), (9), and (10), respectively.
Note that there are also alternatively adopted subdwarf types given
in column (10) for nine of the objects initially classified as M3-M6 
(see section~\ref{sptsdM}).

   \begin{figure}
   \centering
   \includegraphics[width=90mm,angle=270]{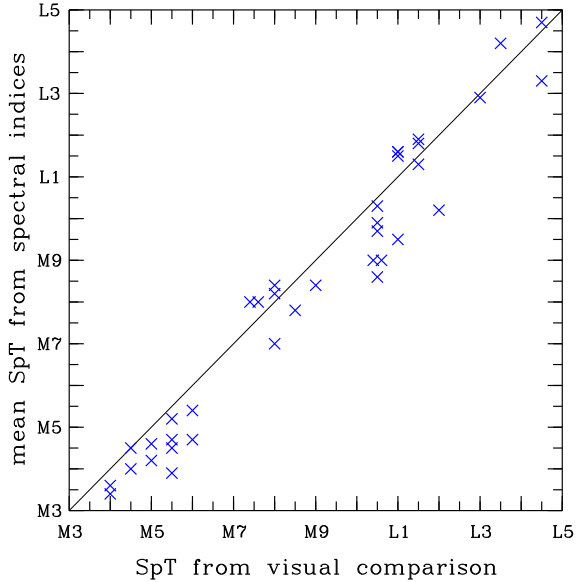}
      \caption{Mean spectral types obtained from spectral indices
vs. spectral types obtained from visual comparison with M and L
dwarf template spectra.
All our targets (except the CWDs and the peculiar object 
SDSS J021546.76$+$010019.2) are shown together with the comparison L
dwarfs (except two late-L dwarfs). For four objects, the visual spectral
types were shifted by $\pm$0.1 subclasses for better visibility.
The line indicates consistence of spectral types.
              }
         \label{fig_sptypes}
   \end{figure}

The relatively large errors of the individual spectral indices and the
corresponding spectral types, as well as the whole procedure for determining 
spectral types 
(as a weighted mean of index-based and visual results)
were in general similar to the errors and the method used in 
Phan-Bao et al.~(\cite{phanbao08}). However, there are some differences:
1) We used the formula for L1-L6 dwarfs from Mart\'{\i}n
et al.~(\cite{martin99}) when the PC3 index was larger than 2.5,
as was also done by Mart\'{\i}n et al.~(\cite{martin06}).
2) For the final spectral types, we took the average of two values, of
the spectral type determined on the one hand from visual comparison 
with template spectra, and of the mean spectral type obtained 
on the other hand from the three (PC3, TiO5, VOa) spectral indices. 
The uncertainty of the finally adopted spectral types is 
conservatively estimated as one spectral type.

In case of the new L4 dwarf SDSS J011014.30$+$010619.1
we did not use the 
spectral type corresponding to the TiO5 index in the computation of the
finally adopted spectral type. 
This object has not only one of the largest uncertainties
in its TiO5 index, but the index is also much larger than the upper
limit (1.5) shown in the spectral index/spectral type relation of
Cruz \& Reid~(\cite{cruz02}; their Fig.3).
For the two comparison late-L dwarfs, we have not used the 
average of index-based spectral types, as the
Cruz \& Reid~(\cite{cruz02}) spectral type/spectral index relations
are only valid up to mid-L types.
For all our comparison objects we adopted 
the previously known optical spectral types (Tab.~\ref{tabLTpm}), 
except for our latest-type object SDSSp J023617.93$+$004855.0,
which we re-classified based on the comparison with the spectrum of the
L8 dwarf SDSSp J010752.33$+$004156.1 (Fig.~\ref{spec_L}) as an L9 dwarf
(see also Sect.~\ref{sec4}). The spectrum of 2MASS J00521232$+$0012172,
for which only a near-infrared spectral type was known before 
(Tab.~\ref{tabLTpm}), is very similar to that of the L4.5 dwarf 
2MASSI J0104075$-$005328 and was also classified as L4.5 based on 
its spectral indices.

We have compared the results from the visual spectral classification
using normal M and L dwarf templates
with those based on the spectral indices (Fig.~\ref{fig_sptypes}) and found 
small systematic differences. For all 37 objects investigated, we found a 
mean difference (visual minus index-based) of $+$0.46 (with a standard 
deviation of 0.76) subtypes. A larger difference of $+$0.69 (st. dev. 0.44)
is measured for 12 M3-M6 dwarfs, whereas the effect is smaller for the
25 M7-L5 dwarfs: $+$0.35 (st. dev. 0.85). The largest individual differences
reach 1.6 and 1.9 subtypes, respectively for the M3-M6 and M7-L5 dwarfs.
We note that the PC3 index alone gives already very similar
results to our visual classification. The two other indices show a much
larger scatter and different systematic deviations, which however 
compensate each other in the averaging. 
The small systematic errors and standard deviations given above are
well below our conservatively estimated classification accuracy of one
spectral type and show that the two methods (index-based and visual) 
lead to similar results.

In total, we classified 13 new L dwarfs and assigned new spectral types
to two previously 
known L dwarfs from our optical spectra. We also found 8 new late-type
M dwarfs and 13 mid-type M dwarfs. In the following we have searched 
for subdwarf candidates among the M-type objects.

\subsection{M (sub)dwarfs}
\label{sptsdM}

From the visual comparison of the new target spectra with the 
comparison cool subdwarf spectra shown in the lower left part of
Fig.~\ref{spec_cwdsdM} we did not find any obvious similarities.
In particular, none of the late-M and L-type spectra resembled
one of the three ultracool ($>$sdM7) subdwarfs. 
However, among the mid-M dwarfs, shown in more detail in
Fig.~\ref{plot_MsdM_1} and Fig.~\ref{plot_MsdM_2}, there are many objects
which can be also fitted by the sdM7 template spectrum of LHS~377.  
This is due to our noisy spectra
where the small but significant differences between the sdM7 
(Gizis~\cite{gizis97}) and the M4.5-M5.5 
(Kirkpatrick et al.~\cite{kirkpatrick99}) templates are hardly
seen. The spectrum of SDSS J024958.88$+$010624.3 fits nearly
perfectly with that of LHS~377, whereas two others 
(SDSS J215515.49$+$005128.0 and SDSS J231914.40$+$005615.9) 
are very close to that. We assign an alternative spectral type of
sdM7 to these three targets. For four others (re-classified as sdM7:) 
we can still see a reasonable agreement with the spectrum of LHS~377.
However, none of the sdM7-like spectra comes close to that of the
sdM9.5 comparison object SSSPM J1013$-$1356 shown at the bottom of
Fig.~\ref{plot_MsdM_2}.
Our earliest type M dwarf spectrum (SDSS J221911.35$+$010220.5;
Fig.~\ref{plot_MsdM_1}) fits reasonably well to
that of LHS~320 (sdM2; Gizis~\cite{gizis97}).

\begin{figure}
   \centering
   \includegraphics[width=90mm]{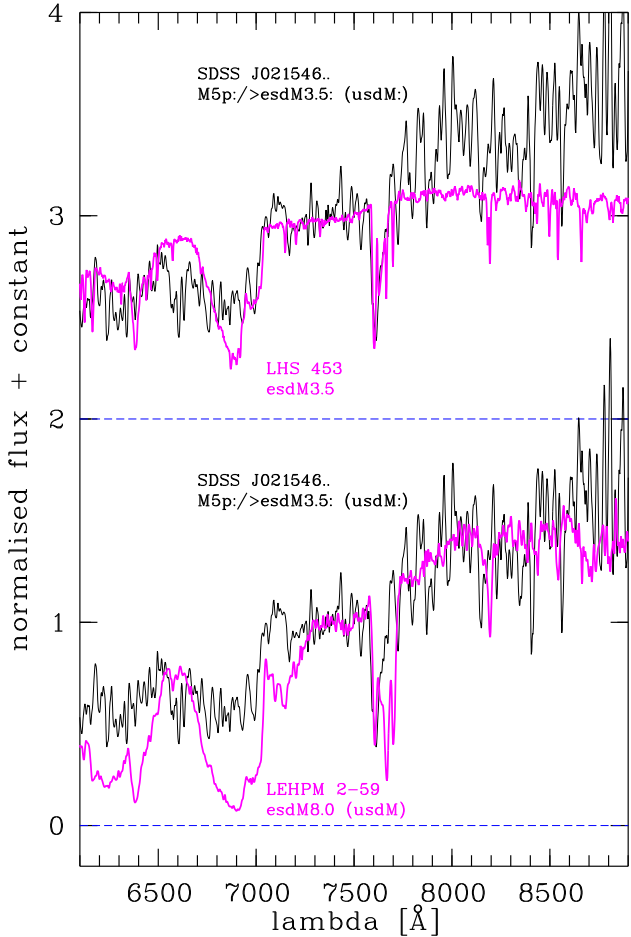}
      \caption{Spectrum of the peculiar M5p:/$>$esdM3.5:
               object SDSS J021546.76$+$010019.2 (usdM: according
               to the new system of L{\'e}pine et al.~\cite{lepine07})
               compared to that of the comparison esdM8.0 (usdM) object
               LEHPM 2-59 (both spectra taken with VLT/FORS1) and
               to the Keck/LRIS spectrum of LHS~453 (esdM3.5; 
               Gizis~\cite{gizis97})
               from Neill Reid's webpage. The comparison spectra
               (thin lines) were not smoothed, whereas the noisy
               spectrum of SDSS J021546.76$+$010019.2 was smoothed with a
               running mean over 7 pixels (thick line).
              }
         \label{fig_esdM}
   \end{figure}

The spectrum of SDSS J021546.76$+$010019.2 is a peculiar one.
On the red side, the continuum is as strong as that of an M5 dwarf
(Fig.~\ref{plot_MsdM_1}), whereas on the blue side the signal is
much stronger but relatively featureless. We speculate that this
may be caused by an unresolved white dwarf companion (for comparison
see e.g. Raymond et al.~\cite{raymond03}). It 
looks also like one of the mid-M extreme 
subdwarfs (esdM in the system of Gizis~\cite{gizis97}), which we
extracted from the above mentioned web page of Neill Reid, while
its continuum is as red as that of the known esdM8 object
LEHPM 2-59 
(see Fig.~\ref{fig_esdM})
but does not show 
either 
the characteristic absorption 
bands bluewards of 7000\AA\, 
or 
the strong KI doublet at 7665/7699\AA\,.
Unfortunately, this peculiar spectrum has only very low
signal-to-noise so that we assign an uncertain type of M5p:/$>$esdM3.5:.
For all nine objects with an alternative subdwarf classification
(Tab.~\ref{tabsptdv}), we estimate an uncertainty of at least
one subdwarf type.

%
\begin{table*}
\caption{Additional $ugr$ photometry of objects with well-measured $gr$ magnitudes (for $iz$ magnitudes see Tab.~\ref{tabnewhpm})}
\label{tabugr}
\centering
\begin{tabular}{l c c c c c c c c r r}
\hline\hline
Name from SDSS     &  SpT          & mean $u$ & mean $g$ & mean $r$ & $u-g$   & $g-r$   & $r-i$   & $i-z$   & $N_{gr}$ & $N_{u}$ \\
DR6 (SDSS~J...)    &               & [mag]    & [mag]    & [mag]    & [mag]   & [mag]   & [mag]   & [mag]   &          &         \\
(1)                & (2)           &  (3)     & (4)      & (5)      & (6)     & (7)     & (8)     & (9)     & (10)     & (11)    \\
\hline
010959.91$-$010156.3 & CWD   & $>$23.987 & 23.420 & 22.346 & $>$$+$0.567 & $+$1.074 & $+$0.361 & $+$0.261 & 23 & 14 \\
   & &~~$\pm$0.275 &$\pm$0.080 &$\pm$0.120 &  ~~$\pm$0.286&$\pm$0.144&$\pm$0.137&$\pm$0.135\\
020508.04$+$002458.7 & CWD   & $>$24.114 & 22.216 & 21.532 & $>$$+$1.899 & $+$0.684 & $+$0.276 & $+$0.045 & 44 & 35 \\
   & &~~$\pm$0.232 &$\pm$0.020 &$\pm$0.013 &  ~~$\pm$0.233&$\pm$0.024&$\pm$0.026&$\pm$0.076\\
032129.20$+$003212.8 & CWD   & $>$24.684 & 22.663 & 21.765 & $>$$+$2.021 & $+$0.899 & $+$0.319 & $+$0.141 & 21 & 15 \\
   & &~~$\pm$0.362 &$\pm$0.054 &$\pm$0.021 &  ~~$\pm$0.366&$\pm$0.060&$\pm$0.033&$\pm$0.096\\
222939.14$+$010405.5 & CWD   & $>$24.461 & 22.609 & 21.892 & $>$$+$1.851 & $+$0.718 & $+$0.345 & $-$0.341 & 15 & 12 \\
   & &~~$\pm$0.443 &$\pm$0.045 &$\pm$0.040 &  ~~$\pm$0.445&$\pm$0.060&$\pm$0.067&$\pm$0.178\\
\\
221911.35$+$010220.5 & M3.5/sdM2:  & $>$24.976 & 24.267 & 22.654 & $>$$+$0.709 & $+$1.613 & $+$0.988 & $+$0.325 &  9 &  4 \\
   & (sdM:)&~~$\pm$0.660 &$\pm$0.273 &$\pm$0.098 & ~~$\pm$0.714&$\pm$0.290&$\pm$0.103&$\pm$0.144 \\
\hline
\end{tabular}
\end{table*}

Although most of our spectra have only low signal-to-noise, which combined
with the low resolution do not allow accurate measurements of the CaH
and TiO5
indices used in the classification of M subdwarfs (Gizis~\cite{gizis97}),
we list the CaH1, CaH2, and CaH3 indices together with their errors
in Tab.~\ref{tabspind}.
Note that the accuracy of the CaH indices and of the TiO5 index
in Gizis~(\cite{gizis97}) was $\pm$0.02-0.04. This accuracy has been
reached by us only in exceptional cases for all four indices.
In the last column
of Tab.~\ref{tabspind} we give the $\zeta_{TiO/CaH}$ defined by
L{\'e}pine et al.~(\cite{lepine07}) for creating a new three-class
(sdM, esdM, usdM) system instead of the two-class (sdM, esdM) system
of Gizis~(\cite{gizis97}). The $\zeta_{TiO/CaH}$ depend only
on TiO5, CaH2, and CaH3. In case of our spectra, the latter three
indices are generally better defined than the CaH1 index. We list the
subdwarf class according to the $\zeta_{TiO/CaH}$ values for
all M-type objects in Tab.~\ref{tabsptdv}. There are seven moderately
metal-poor subdwarf (sdM) candidates and one very low metallicity (usdM:)
candidate, which is the already mentioned above esdM-like object
SDSS J021546.76$+$010019.2, among the M dwarfs in our sample.
Note that the esdM8.0 comparison object LEHPM 2-59 is also an usdM
according to the new scheme of L{\'e}pine et al.~(\cite{lepine07}).
The two new M8.0 and M8.5 dwarfs with an sdM: indication from 
their $\zeta_{TiO/CaH}$, SDSS J003435.32$+$004633.9 and 
SDSS J005837.48$+$004435.6, show spectra clearly distinctive 
from both late-subdwarf  (sdM9.5 and esdM8.0) comparison spectra 
(bottom of Fig.~\ref{spec_cwdsdM}).

With 9 out of 13 objects with an alternative subdwarf 
classification and two more objects with a hint on a subdwarf 
nature obtained from their $\zeta_{TiO/CaH}$ it seems possible 
that all the mid-M dwarfs are in fact subdwarfs. However, there 
is one object, SDSS J013415.85$+$001456.0, classified as M5.5 
and possibly being an sdM: according to its $\zeta_{TiO/CaH}$, 
which has a spectrum clearly typical of a normal mid-M dwarf 
(Fig.~\ref{plot_MsdM_2}).  
In the spectrum of SDSS J011755.09$+$005220.0, reclassified as 
sdM7: and having an sdM: flag from its $\zeta_{TiO/CaH}$, we may 
see an H$\alpha$ line in emission (top of Fig.~\ref{plot_MsdM_2}),
which is usually not observed in subdwarfs and would support its
initial classification as a normal M5 dwarf. 

\subsection{cool white dwarfs}
\label{sptcwd}

Four new HPM objects (all with $i-z<+0.4$)
show featureless spectra similar to that of the known cool 
white dwarf SDSSp J133739.40$+$000142.8 (upper left part of 
Fig.~\ref{spec_cwdsdM}). Compared to the latter the new
objects show a less pronounced blue continuum in their spectra.
However, they are much bluer than the earliest M (sub)dwarfs
in our sample, and their $gr$ magnitudes are well measured.
We list the additional photometry for those four objects in
Tab.~\ref{tabugr} together with that of the only M 
(sub)dwarf
with well-measured $gr$ magnitudes, SDSS J221911.35$+$010220.5
(M3.5/sdM2:),
which is the fifth object with $i-z<+0.4$ similar to the
four CWDs but with an uncertain proper motion
(see Fig.~\ref{fig1}). The $u$ magnitudes given
in Tab.~\ref{tabugr} are lower limits, since all our faint
($i>21$) objects were only partly detected in that passband
(the number of detections in $gr$ and in $u$ are given in the 
last two columns, respectively). 
As one can see, not only the spectrum of SDSS J221911.35$+$010220.5
is clearly distinctive but also its $g$$-$$r$ and $r$$-$$i$ colours are
much redder than those of the CWDs.

Our spectra cover a smaller wavelength interval than
used for the classification of other CWDs 
discovered in the SDSS (e.g. Gates et al.~\cite{gates04}; 
Harris et al.~\cite{harris08}), and more spectroscopic data
(and model atmosphere fitting) are needed to better characterise 
our objects. However, the accurately measured large
proper motions and colours (Tab.~\ref{tabnewhpm}, Tab.~\ref{tabugr})
of the four objects with featureless
spectra support their classification as CWDs.
Their $g-r$, $r-i$ and $i-z$ colours are similar to those of
other CWDs with mild collision-induced absorption
(CIA) described by Harris et al.~(\cite{harris08}). On the other
hand, the only observed by us comparison CWD 
SDSSp J133739.40$+$000142.8 is typical of CWDs
with strong CIA suppression and temperatures below 4000~K 
(Gates et al.~\cite{gates04}).
Our new CWDs are probably warmer.

%

\section{Distance estimates and kinematics}
\label{sec4}

Spectroscopic distances of all M and L dwarfs and their resulting
tangential velocities (Tab.~\ref{tabsptdv})
were derived using absolute magnitudes
$M_z$ of L dwarfs later than L2 given by Hawley et al.~(\cite{hawley02})
and the revised data with updated SDSS colours for M0 to L2 dwarfs
from West et al.~(\cite{west05}). The values for half-step spectral
types were determined by linear interpolation between the neighbouring
absolute magnitudes for integer spectral types.

For an estimate of the expected uncertainties in the distances
and tangential velocities, we have made a conservative assumption of
0.5~mag accuracy in the absolute magnitudes of all late-M and L dwarfs. 
This includes an uncertainty of one spectral type corresponding to about
0.3-0.4~mag differences in $M_z$. For M3-M6 dwarfs, the uncertainty of
one spectral type corresponds to 0.8-1.2~mag differences in $M_z$. Therefore,
we use a 1.0~mag accuracy in the absolute magnitudes of M3-M6 dwarfs for
our distance calculation. Note that for all M and L dwarfs the uncertainties
in colours are relatively small (0.1-0.2~mag), and that the changes in
$z-J$ in the recent paper by West et al.~(\cite{west08}) compared to
those in West et al.~(\cite{west05}) are also of that order. Our assumed
uncertainties in the absolute magnitudes lead to 23\% and 46\% distance
(and tangential velocity) errors, respectively for M7-L9 and M3-M6 dwarfs.
The errors in the proper motions have a much smaller effect on the
uncertainties of the tangential velocities and were neglected. However,
for objects with uncertain proper motions marked in 
Tab.~\ref{tabnewhpm} and Tab.~\ref{tabsptdv} this is not the case. Their 
tangential velocities may be overestimated by an unknown factor 
corresponding to the possibly overestimated proper motions.

For a second distance estimate of the nine mid-M dwarfs 
with alternatively adopted subdwarf types, we have used the 
trigonometric parallaxes of the corresponding comparison objects,
LHS~453 and LHS~377 (10.3$\pm$0.9~mas and 28.4$\pm$0.8~mas, 
respectively from Monet et al.~\cite{monet92}), and  
LHS~320 (25.8$\pm$3.6~mas from van Altena et 
al.~\cite{vanaltena95}).  All three comparison objects have 
been measured in the SDSS. We have used the $z$ magnitudes 
from SDSS DR6 (Adelman-McCarthy et al.~\cite{adelman08}) of 
12.878$\pm$0.006 and 14.673$\pm$0.006, respectively for 
LHS~320 and LHS~377.  For LHS~453, we used the SDSS DR7 
(Abazajian et al.~\cite{abazajian08}) $z$ magnitude of 
15.897$\pm$0.007. The absolute magnitudes $M_z$ are
9.94$\pm$0.30, 10.96$\pm$0.19, and 11.94$\pm$0.06, respectively 
for LHS~320 (sdM2), LHS~453 (esdM3.5), and LHS~377 (sdM7).  
For the resulting alternative distances and tangential 
velocities of our targets (Tab.~\ref{tabsptdv}) we estimate 
again errors of about 46\% based on our assumed larger 
uncertainty of 1~mag in the absolute magnitudes which is 
dominated by the expected uncertainty of more than one 
type in our subdwarf classification.

   \begin{figure}
   \centering
   \includegraphics[width=110mm]{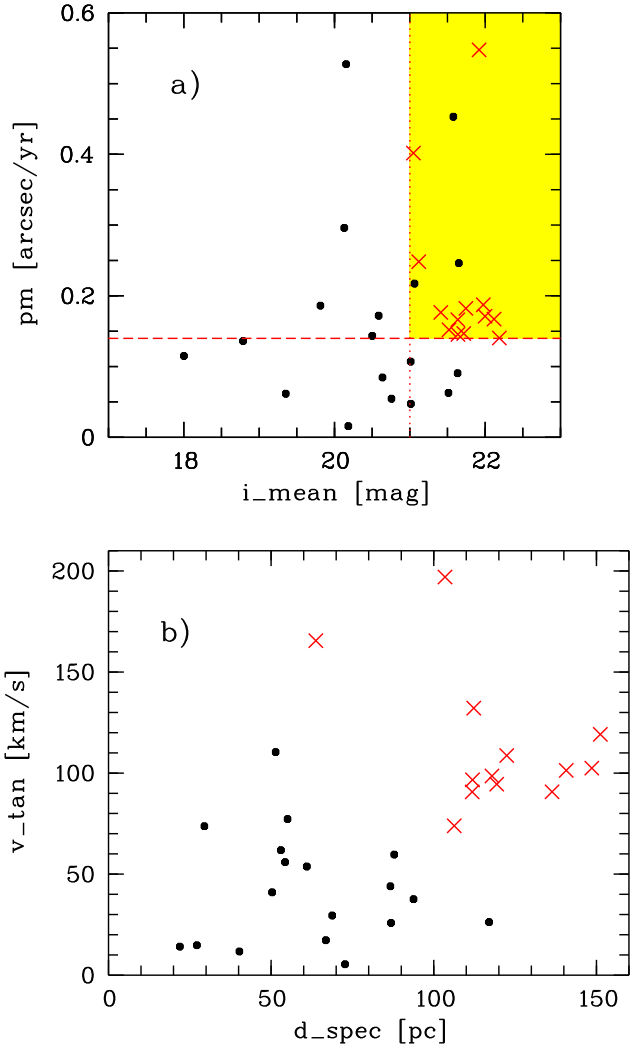}
      \caption{{\bf a)} Mean $i$ magnitudes and proper motions
of all previously known (dots) and new (crosses) L0-L4.5 dwarfs in S82.
The area right of the dotted line and above the dashed line marks
the selection criteria of our proper motion survey.
{\bf b)} Spectroscopic distances and tangential velocities
of previously known (dots) and new (crosses) L0-L4.5 dwarfs in S82.
              }
         \label{fig_Ldvtan}
   \end{figure}

Since all newly discovered L dwarfs have spectral types between L0 and L4.5
we compare their measured mean $i$ magnitudes and proper motions as well as
the determined spectroscopic distances and tangential velocities with the
data of all previously known L0-L4.5 dwarfs in S82 in Fig.~\ref{fig_Ldvtan}.
As one can see, the known early-L dwarfs are mostly brighter and have on
the average smaller proper motions. Consequently, the new L dwarfs have
larger distances (between 60 and 150 pc) and tangential velocities (70 to
200 km/s), which are typical of the Galactic thick disk and halo. None of
the new L dwarfs has been marked as an uncertain proper motion object
($^{\dagger}$ in Tab.~\ref{tabnewhpm} and Tab.~\ref{tabsptdv}),
so that their relatively large velocities seem to be reliable. There are 
two new L dwarfs (SDSS J005636.22$-$011131.9 and SDSS J235841.98$+$000622.0) 
with formal proper motion errors larger than $\pm$25~mas/yr, but their 
tangential velocities are only moderately large ($\sim$100~km/s).
With this study we have considerably increased the number of 
high-velocity L dwarfs compared to follow-up proper motion determinations
of large numbers of L dwarfs (Schmidt et al.~\cite{schmidt07};
Faherty et al.~\cite{faherty08}) discovered previously in large area 
(all-sky) surveys.
Among a total of more than 400 late-M and L dwarfs investigated, the latter
two studies include only 10 normal L dwarfs with $v_{tan}$$>$100~km/s, whereas
we have discovered 7 in a much smaller sky area. This underlines that
a HPM survey which goes to fainter magnitudes is a powerful tool
to detect those ''fast'' L dwarfs.

The spectroscopic distance of the L8 dwarf SDSSp J010752.33$+$004156.1
(17$\pm$4~pc) 
agrees very well with its trigonometric parallax of 
64.13$\pm$4.51~mas measured by Vrba et al. (\cite{vrba04}).
This agreement supports our re-classification of the previously known
L6 dwarf SDSSp J023617.93$+$004855.0 as an L9 dwarf (Sect.~\ref{sptML}).
An alternative earlier (L5.5) classification of SDSSp J010752.33$+$004156.1
while accepting the L6 type of SDSSp J023617.93$+$004855.0 would lead to
a much larger spectroscopic distance 
(27$\pm$6~pc) 
of SDSSp J010752.33$+$004156.1,
i.e. nearly two times larger than the trigonometric distance estimate.

Even larger distances (180 to 280 pc) and corresponding tangential velocities
(130 to 320 km/s) have been estimated for the eight late-M dwarfs in our 
sample. The largest velocity corresponds to the only uncertain proper 
motion object (SDSS J225903.29$-$004154.2) among them. Three objects
(SDSS J215934.25$+$005308.5, SDSS J232935.99$-$011215.3, 
SDSS J235826.48$+$003226.9) have proper motion errors larger 
than $\pm$25~mas/yr, and two others are mentioned as sdM candidates. 
Despite these uncertainties, the late-M dwarfs belong probably
also to the thick disk/halo populations.
None of the seven late-M dwarfs with tangential velocities 
of the order of 200~km/s and more, similar to those of the three comparison
ultracool subdwarfs (Tab.~\ref{tabsptdv}), 
show obvious low-metallicity indications
in their spectra, i.e. they do not resemble the spectra of any 
of these ultracool subdwarfs (Fig.~\ref{spec_cwdsdM}).
According to our findings, there is apparently a population
of high-velocity normal late-M dwarfs in addition to that of 
ultracool subdwarfs.  Note that Faherty et
al.~(\cite{faherty08}) have found only three normal late-M dwarfs
with tangential velocities exceeding 100~km/s among $>$250 previously
classified M7-M9.5 dwarfs with lacking proper motions, whereas 
Schmidt et al.~(\cite{schmidt07}) have failed to find any among
about 80 investigated M7-M9.5 dwarfs within 20~pc.

The spectroscopic distances of the 13 mid-M dwarfs range from 
about 600 to 2500 pc, and their tangential velocities lie between 
about 400 and 2500 km/s, if the proper motions are correct. However,
seven out of nine objects with tangential velocities above 700 km/s
are uncertain proper motion objects (Tab.~\ref{tabsptdv}) so that 
we consider these very large velocities as highly uncertain, too.
The majority of the mid-M dwarfs have an alternative subdwarf
classification, and we can not exclude that all the mid-M dwarfs
with rather noisy spectra are in fact subdwarfs. The subdwarf 
classification reduces our extreme distance and tangential velocity 
estimates by up to about 50\% (see also Scholz et 
al.~\cite{scholz05}). Among the four mid-M (sub)dwarfs not
marked as uncertain proper motion objects, there are the 
M5p:/$>$esdM3.5: object 
SDSS J021546.76$+$010019.2 and the 
M4.5 dwarf SDSS J025316.06$+$005157.1 with velocities of the order 
of 1000~km/s. Their proper motion errors are larger than 
$\pm$25~mas/yr and amount to about 20-25\% uncertainty in their
proper motion values. Our assumed total uncertainty of 46\% in their
tangential velocities, which is dominated by the uncertain spectral 
types and absolute magnitudes, is still consistent with a velocity
below the local Galactic escape velocity (Smith et al.~\cite{smith07}).
For many of
the 13 objects classified as mid-M dwarfs our proper motions
are probably overestimated.
But those, for which the proper motion measurement
is approximately correct, are likely Galactic halo members.

   \begin{figure}
   \centering
   \includegraphics[width=110mm]{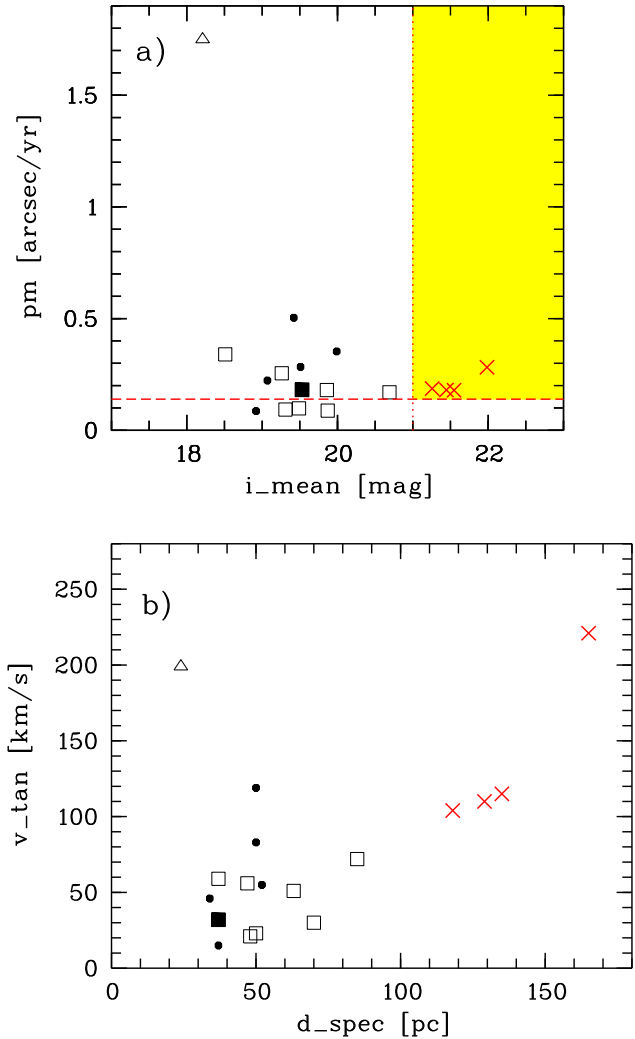}
      \caption{{\bf a)} Mean $i$ magnitudes and proper motions
of new CWDs discovered in S82 (crosses) compared to those of
other spectroscopically confirmed CWDs discovered from SDSS:
filled square - Harris et al.~(\cite{harris01}),
dots - Gates et al~(\cite{gates04}),
open squares - Harris et al.~(\cite{harris08}),
triangle - Hall et al.~(\cite{hall08}).
The area right of the dotted line and above the dashed line marks
the selection criteria of our S82 proper motion survey.
{\bf b)} Spectroscopic distances and tangential velocities
of the new (crosses) and previously discovered SDSS CWDs
(symbols as in a) with same objects shown). Mean values
of the different estimates for all previously known CWDs 
were used here. 
              }
         \label{fig_CWDdvtan}
   \end{figure}

As already mentioned in Sect.~\ref{sptcwd}, our four new CWDs have
SDSS colours consistent with the mild-CIA class of CWDs described 
in Harris et al.~(\cite{harris08}). In order to get distance estimates
for our targets we have used two of the objects discovered by Harris et 
al.~(\cite{harris08}), SDSS J0310$-$01 and SDSS J1632$+$24 (=LP386-28;
Luyten~\cite{luyten79b}) for comparison. The latter two objects have
according to the $i$ magnitudes and distances given by Harris et al.
(19.87, 40-99~pc and 18.51, 21-52~pc, respectively for SDSS J0310$-$01 
and SDSS J1632$+$24) very similar absolute magnitudes of $M_i=15.9$
with an uncertainty of 1.0 mag. We have assigned this mean absolute magnitude
to all our CWDs and determined distances between about 120 and 170~pc.
The assumed uncertainty of 1.0~mag in the absolute magnitudes
corresponds to 46\% errors in the distances and tangential velocities
of the new CWDs. The proper motion errors of all new CWDs are small so
that we have neglected their effect in the uncertainties of the computed
tangential velocities.
Three of the resulting tangential velocities lie between 100 and 120~km/s
and indicate thick disk or halo membership, whereas the faintest CWD, 
SDSS J010959.91$-$010156.3 has a larger tangential velocity of about
220~km/s which qualifies this object as a likely halo member.

Compared to spectroscopically confirmed CWDs formerly detected in
SDSS data (Harris et al.~\cite{harris01}; Gates et al~\cite{gates04};
Harris et al.~\cite{harris08}; Hall et al.~\cite{hall08})
our new CWDs are generally fainter by about 2 mag but still
have similarly large (and well-measured) proper motions (Fig.~\ref{fig_CWDdvtan}a). 
Note, there is an important difference between the last two figures:
Whereas the comparison early-L dwarfs in
Fig.~\ref{fig_Ldvtan} represent the previously known census
in the S82 area, the comparison CWDs in Fig.~\ref{fig_CWDdvtan} 
have been detected in the whole SDSS, i.e. in a sky area about 
40 times larger than our survey area.
Since we assume our new CWDs to have the same mean absolute 
magnitudes as the two comparison objects from Harris et 
al.~(\cite{harris08}; see above), the new objects 
are consequently at much larger distances 
and have exclusively thick disk and halo kinematics,
whereas most of the previous SDSS discoveries have smaller velocities
typical of the thin disk population (Fig.~\ref{fig_CWDdvtan}b). However,
we can not exclude that some of our new CWDs have fainter absolute 
magnitudes. Trigonometric parallax measurements are needed 
to clarify this issue.

%

\section{Conclusions}
\label{sec5}

   \begin{enumerate}
      \item We have carried out a HPM survey of faint objects in S82
            using their individual epoch positions from 1998 to 2004.
            The 38 faintest HPM objects ($i$$>$21, $\mu$$>$0.14~arcsec/yr)
            have been selected (together with some comparison objects) for our
            spectroscopic follow-up observations. Very recently,
            in an independent attempt
            to find moving sources in S82, Lang et al.~(\cite{lang08}) used
            a method of fitting a model of a moving point source to all
            S82 imaging data. Their initial candidate list was selected
            from the S82 ``Co-add catalog'' (J. Annis et al., in prep.)
            aiming at the detection of 
            HPM objects that are too faint to detect at any individual epoch.
            Among 
            their 19 HPM brown dwarf candidates (with proper motions
            between 0.07 and 0.65~arcsec/yr)
            there are 10 previously
            known objects (nine L dwarfs and one T dwarf, which were all
            detected in our survey, too. Among the remaining objects of
            Lang et al. there are only four with proper 
            motions $\mu>0.14$~arcsec/yr. Two of the 
            latter (SDSS J011014.30$+$010619.1 and
            SDSS J234841.38$-$004022.1) were also on our target list
            and classified as L4 and L1.5, respectively. We conclude that
            most of their objects are still detectable at individual epochs
            as we used in our HPM survey.
      \item The low-resolution (and relatively low signal-to-noise) 
            spectra of our faint targets observed with FORS1@VLT
            allowed us to classify them as early-L (13 objects) , late-M (8),
            mid-M (13) dwarfs, and CWDs (4). In addition, two previously known
            L dwarfs were reclassified by us. None of the 38 spectra looked
            like those of the three ultracool ($>$sdM7) subdwarf spectra
            observed by us for comparison, 
            but 9 out of 13 mid-M dwarfs were alternatively
            classified as subdwarfs (sdM7 and earlier types) from a
            comparison with additional subdwarf templates.
            Relatively uncertain spectral indices measurements 
            do also
            hint at a possible 
            subdwarf nature of
            seven M dwarfs. One mid-M dwarf is likely a very low-metallicity
            (usdM) object
            or may have an unresolved white dwarf companion.
      \item The number of L dwarfs in the S82 area, which was already before
            our study one of the best investigated fields in the sky, has been
            further increased (from 22 to 35). The newly added L dwarfs
            are preferentially thick disk and halo objects as are the new
            late-M dwarfs and CWDs found in S82. The nearest new object is an
            L4 dwarf (SDSS J011014.30$+$010619.1) at 64~pc. All other new M and L
            dwarfs have spectroscopic distance estimates above 100~pc. 
      \item The re-discovery of all formerly known L dwarfs and even of some T 
            dwarfs in S82, including the L8 dwarf SDSSp J010752.33$+$004156.1 and 
            the T4.5 dwarf SDSS J020742.48$+$000056.2
            with trigonometric parallaxes (Vrba et al.~\cite{vrba04})
            placing them at 16 and 29~pc, respectively, shows that our HPM
            survey was sensitive to discover very nearby objects, too.
            {\it However, we failed to detect additional very nearby objects.}
            For two previously known objects reclassified by us,
            the L4.5 dwarf 2MASS J00521232$+$0012172 and the L9 dwarf
            SDSSp J023617.93$+$004855.0, we have determined new spectroscopic
            distances of 53~pc (formerly 81~pc) and 18~pc (formerly 21~pc), respectively.
      \item On the one hand, we may have been just unlucky in finding new L-type
            neighbours of the Sun since our survey covered only 1/150 part of 
            the sky and the total number of missing L dwarfs in the Solar
            neighbourhood (within about 50~pc) may be of the order of several hundreds.
            On the other hand, the S82 area had already been well-investigated
            for the apparently brighter and consequently more nearby L dwarfs.
            Using the L and T dwarf compendium by Gelino et al.~(\cite{gelino08}) and
            the $M_J$/spectral type relation given in Dahn et al.~(\cite{dahn02}),
            we have estimated that about 400 of the known L dwarfs have spectroscopic
            distances of less than 50~pc. 11 of them (seven L0-L4.5 and four L5.5-L9.5
            dwarfs) lie in the S82 area, which corresponds to 
            a 4 times higher surface density than the average. Therefore,
            we conclude that the census of nearby L dwarfs in S82 was
            probably 
            already near-complete 
            before our study. 
            The space density of L dwarfs determined from the 11 
            objects within 50~pc falling in the S82 area is about 
            0.003~pc$^{-3}$. This is comparable with the value we get
            from the probably more
            complete list of about 200 L dwarfs with spectroscopic
            distances within 25~pc in Gelino et al.~(\cite{gelino08}), 
            again using the Dahn et al.~(\cite{dahn02}) $M_J$/spectral
            type relation.
      \item With our deep HPM survey we have discovered 
            preferentially high-velocity late-M and L dwarfs. However,
            the optical spectra of these objects do not show indications
            of a lower metallicity typical of ultracool subdwarfs. As long
            as we have no infrared data on the new fast late-M and L dwarfs, 
            we can not
            say whether they are unusually blue (according to specific
            near-infrared spectral features) or blue photometric outliers
            (acording to their $J-K_s$ colours) (cf. 
            Faherty et al.~\cite{faherty08}). Our findings suggest that
            there may exist a population of normal-metallicity
            ultracool halo objects.
      \item 10 of the 38 HPM objects have uncertain proper motions.
            Most of these uncertain proper motion objects turned out to be
            mid-M dwarfs at large distances. 
            Most of them have been alternatively classified as
            subdwarfs or can be considered as subdwarf candidates, which 
            may in fact have large velocities. However,
            the extremely
            large tangential velocities (700...2500~km/s) of nine objects
            suggest an overestimation of their proper motions.
            A more careful investigation of the kinematics of
            these sources makes sense only after a verification of their
            proper motions with additional epoch measurements. When the
            large proper motions will be confirmed we would also need
            higher resolution and signal-to-noise spectra to investigate
            their metallicities and radial velocities.
      \item Follow-up observations 
            (better spectroscopy and near-infrared photometry
            allowing a fit of the observed spectral energy distribution 
            with CWD atmosphere models for determining the effective
            temperatures)
            are needed to better characterise the four new CWDs.
            Trigonometric parallax measurements are challenging in
            view of their faint magnitudes and possible large distances,
            but would provide the ultimate values of their absolute 
            magnitudes. Their estimated distances of more than 100~pc
            are based on the assumption that their absolute magnitudes
            compare with other CWDs detected in the SDSS.
      \item If we consider the optimistic case that all four new CWDs 
            are Galactic halo members within 170~pc, then we determine a
            space density of 2.9 $\times$ 10$^{-5}$~pc$^{-3}$ {\it for halo CWDs
            alone}, which is almost exactly the same number as the 
            density of 3.0 $\times$ 10$^{-5}$~pc$^{-3}$ derived 
            by Gates et al.~(\cite{gates04}) for a mix of
            {\it all} CWDs in various Galactic components. Even with the
            conservative assumtion that only two of our CWDs are representatives
            of the halo population the estimated space density of halo CWDs 
            reaches already 50\% of the above cited value. Nevertheless,
            our derived space density of halo CWDs is consistent with
            other estimates indicating that ancient CWDs do not contribute
            significantly to the baryonic fraction of the dark halo of the
            Galaxy (see Carollo et al.~\cite{carollo07} and references therein).
            Our estimate of the density of halo CWDs corresponds to
            less than 0.4\% of the local dark matter density 
            (0.0079~$M_{\odot}$pc$^{-3}$) adopted by 
            Torres et al.~(\cite{torres08}).
   \end{enumerate}

\begin{acknowledgements}

We thank the referee, Dr. Adam Burgasser, for his very helpful
detailed comments and suggestions.
We would like to thank Doug Finkbeiner for his
help accessing the SDSS data at Princeton University,
Fernando Comer{\'o}n and the ESO User Support 
Department for their help during phase 2 preparation of the
service mode observations with the VLT, and the 
observers at ESO for carrying out these observations. We also thank
Daniel Bramich, Evalyn Gates, David Hogg and Donald Schneider for helpful 
comments.
Funding for the SDSS and SDSS-II has been provided by the Alfred P. Sloan 
Foundation, the Participating Institutions, the National Science Foundation, 
the U.S. Department of Energy, the National Aeronautics and Space 
Administration, the Japanese Monbukagakusho, the Max Planck Society, and the 
Higher Education Funding Council for England. The SDSS Web Site is 
http://www.sdss.org/.
The SDSS is managed by the Astrophysical Research Consortium for the 
Participating Institutions. The Participating Institutions are the American 
Museum of Natural History, Astrophysical Institute Potsdam, University of 
Basel, University of Cambridge, Case Western Reserve University, University of 
Chicago, Drexel University, Fermilab, the Institute for Advanced Study, the 
Japan Participation Group, Johns Hopkins University, the Joint Institute for 
Nuclear Astrophysics, the Kavli Institute for Particle Astrophysics and 
Cosmology, the Korean Scientist Group, the Chinese Academy of Sciences 
(LAMOST), Los Alamos National Laboratory, the Max-Planck-Institute for 
Astronomy (MPIA), the Max-Planck-Institute for Astrophysics (MPA), New Mexico 
State University, Ohio State University, University of Pittsburgh, University 
of Portsmouth, Princeton University, the United States Naval Observatory, and 
the University of Washington. 
This publication makes use of data products from the Two Micron All Sky Survey, 
which is a joint project of the University of Massachusetts and the Infrared 
Processing and Analysis Center/California Institute of Technology, funded by 
the National Aeronautics and Space Administration and the National Science 
Foundation.

\end{acknowledgements}


\begin{thebibliography}{}

\bibitem[2008]{abazajian08}
Abazajian, K.~N., Adelman-McCarthy, J.~K., Ag\"ueros, M.~A., et al.\ 2008, 
in prep.

\bibitem[2008]{adelman08}
Adelman-McCarthy, J.~K., Ag\"ueros, M.~A., Allam, S.~S., et al.\ 2008,
ApJS, 175, 297

\bibitem[1998]{appenzeller98}
Appenzeller, I., Fricke, K., F\"urtig, W., et al.\ 1998,
Messenger, 94, 1

\bibitem[2006]{artigau06}
Artigau, {\'E}., Doyon, R., Lafreni{\`e}re, D., et al.\ 2006, ApJ, 651, L57

\bibitem[2003]{berriman03}
Berriman, B., Kirkpatrick, D., Hanisch, R., Szalay, A.,
\& Williams, R.\ 2003, IAUJD, 8, 60

\bibitem[2008]{bramich08}
Bramich, D.~M., Vidrih, S., Wyrzykowski, L., et al.\ 2008, MNRAS, 386, 887

\bibitem[2002]{burgasser02}
Burgasser, A.~J., Kirkpatrick, J.~D., Brown, M.~E., et al.\ 2002, ApJ, 564, 421

\bibitem[2003]{burgasser03}
Burgasser, A.~J., Kirkpatrick, J.~D., Burrows, A., et al.\ 2003, ApJ, 592, 1186

\bibitem[2006]{burgasser06a}
Burgasser, A.~J.,
Geballe, T.~R., Leggett, S.~K., Kirkpatrick, J.~D.,
\& Golimowski, D.~A.\ 2006, \apj, 637, 1067

\bibitem[2006]{burgasser06b}
Burgasser, A.~J., \& Kirkpatrick, J.~D.\ 2006, ApJ, 645, 1485

\bibitem[2007]{burgasser07}
Burgasser, A.~J., Cruz, K.~L., \& Kirkpatrick, J.~D.\ 2007, ApJ, 657, 494

\bibitem[2006]{carollo06}
Carollo, D., Bucciarelli, B., Hodgkin, S. T., et al.\ 2006, A\&A, 448, 579

\bibitem[2007]{carollo07}
Carollo, D., Bucciarelli, B., Hodgkin, S.~T., Lattanzi, M.~G., McLean, B.,
Smart, R.~L., \& Spagna, A.\ 2007,
15th European Workshop on White Dwarfs, 372, 113

\bibitem[2008]{chiu08}
Chiu, K., Liu, M.~C., Jiang, L., et al.\ 2008, MNRAS, 385, L53

\bibitem[2006]{cushing06}
Cushing, M.~C., \& Vacca, W.~D.\ 2006, AJ, 131, 1797

\bibitem[2002]{cruz02}
Cruz, K.~L., \& Reid, I.~N.\ 2002, AJ, 123, 2828

\bibitem[2003]{cruz03}
Cruz, K.~L., Reid, I.~N., Liebert, J., et al.\ 2003, AJ, 126, 2421

\bibitem[2002]{dahn02}
Dahn, C.~C., Harris, H.~C., Vrba, F.~J., et al.\ 2002, AJ, 124, 1170

\bibitem[2008]{dahn08}
Dahn, C.~C., Harris, H.~C., Levine, S.~E., et al.\ 2008, ApJ, 686, 548 

\bibitem[1997]{delfosse97}
Delfosse, X., Tinney, C.~G., Forveille, T., et al.\ 1997, A\&A, 327, L25

\bibitem[2008]{delorme08}
Delorme, P., Delfosse, X., Albert, L., et al.\ 2008, A\&A, 482, 961

\bibitem[1997]{epchtein97}
Epchtein, N., de Batz, B., Capoani, L., et al.\ 1997, The Messenger, 87, 27

\bibitem[2008]{faherty08}
Faherty, J.~K.,
Burgasser, A.~J., Cruz, K.~L., Shara, M.~M., Walter, F.~M.,
\& Gelino, C.~R.\ 2008, arXiv:0809.3008

\bibitem[2000]{fan00}
Fan, X., Knapp, G.~R., Strauss, M.~A., et al.\ 2000, AJ, 119, 928

\bibitem[2004]{finkbeiner04}
Finkbeiner, D.~P., Padmanabhan, N., Schlegel, D.~J., et al.\ 2004, AJ, 128, 2577

\bibitem[2003]{flynn03}
Flynn, C., Holopainen, J., \& Holmberg, J.\ 2003, MNRAS, 339, 817

\bibitem[2007]{folkes07}
Folkes, S.~L., Pinfield, D.~J., Kendall, T.~R., \& Jones, H.~R.~A.\ 2007,
MNRAS, 378, 901

\bibitem[2008]{frieman08}
Frieman, J.~A., Bassett, B., Becker, A., et al.\ 2008, AJ, 135, 338

\bibitem[1996]{fukugita96}
Fukugita, M., Ichikawa, T., Gunn, J.~E., et al.\ 1996, AJ, 111, 1748

\bibitem[2004]{gates04}
Gates, E., Gyuk, G., Harris, H.~C., et al.\ 2004, ApJ, 612, L129

\bibitem[2002]{geballe02}
Geballe, T.~R., Knapp, G.~R., Leggett, S.~K., et al.\ 2002, ApJ, 564, 466

\bibitem[2008]{gelino08}
Gelino, C.~R., Kirkpatrick, J.~D., \& Burgasser, A.~J.\ 2008,
online database for 649 L and T dwarfs at DwarfArchives.org
(status: 22 April, 2008)

\bibitem[1997]{gizis97}
Gizis, J.~E.\ 1997, AJ, 113, 806

\bibitem[2002]{gizis02}
Gizis, J.~E., Reid, I.~N., \& Hawley, S.~L.\ 2002, AJ, 123, 3356

\bibitem[2006]{gizis06}
Gizis, J.~E., \& Harvin, J.\ 2006, AJ, 132, 2372

\bibitem[1998]{gunn98}
Gunn, J.~E., Carr, M., Rockosi, C., et al.\ 1998, AJ, 116, 3040

\bibitem[2006]{gunn06}
Gunn, J.~E., Siegmund, W.~A., Mannery, E.~J., et al.\ 2006, AJ, 131, 2332

\bibitem[2008]{hall08}
Hall, P.~B., Kowalski, P.~M., Harris, H.~C., et al.\ 2008, AJ, 136, 76

\bibitem[1997]{hambly97}
Hambly, N.~C., Smartt, S.~J., \& Hodgkin, S.~T.\ 1997, ApJ, 489, L157

\bibitem[2001a]{hambly01a}
Hambly, N.~C.,  MacGillivray, H.~T.,
     Read M.~A., et al.\ 2001a, MNRAS, 326, 1279

\bibitem[2001b]{hambly01b}
Hambly, N.~C., Irwin, M.~J., \& MacGillivray,
     H.~T.\ 2001b, MNRAS, 326, 1295

\bibitem[2001c]{hambly01c}
Hambly, N.~C., Davenhall, A.~C., Irwin, M.~J., \&
     MacGillivray, H.~T.\ 2001c, MNRAS, 326, 1315

\bibitem[2004]{hambly04}
Hambly, N.~C., Henry, T.~J., Subasavage, J.~P., Brown, M.~A., \& Jao, W.-C.\
2004, AJ, 128, 437

\bibitem[1994]{hamuy94}
Hamuy, M., Suntzeff, N.~B., Heathcote, S.~R., et al.\ 1994, PASP, 106, 566

\bibitem[1999]{harris99}
Harris, H.~C., Dahn, C.~C., Vrba, F.~J., et al.\ 1999, ApJ, 524, 1000

\bibitem[2001]{harris01}
Harris, H.~C., Hansen, B.~M.~S., Liebert, J., et al.\ 2001, ApJ, 549, L109

\bibitem[2008]{harris08}
Harris, H.~C., Gates, E., Gyuk, G., et al.\ 2008, ApJ, 679, 697

\bibitem[2002]{hawley02}
Hawley, S. L., Covey, K. R., Knapp, G. R., et al.\ 2002, AJ, 123, 3409

\bibitem[2002]{henry02}
Henry, T.~J., Walkowicz, L.~M., Barto, T.~C.,
     \& Golimowski, D.~A.\ 2002, AJ, 123, 2002

\bibitem[2006]{henry06}
Henry, T.~J., Jao, W.-C., Subasavage, et al.\ 2006, AJ, 132, 2360

\bibitem[1905]{hertzsprung05}
Hertzsprung, E.\ 1905, Zeit. Phot., 3, 429

\bibitem[2001]{hogg01}
Hogg, D.~W., Finkbeiner, D.~P., Schlegel, D.~J., \& Gunn, J.~E.\ 2001,
AJ, 122, 2129

\bibitem[2000]{ibata00}
Ibata, R., Irwin, M., Bienaym{\'e}, O., Scholz, R., \& Guibert, J.\ 2000,
ApJ, 532, L41

\bibitem[2004]{ivezic04}
Ivezi{\'c}, {\v Z}., Lupton, R.~H.; Schlegel, D., et al.\ 2004, AN, 325, 583

\bibitem[2007]{kendall07}
Kendall, T.~R., Jones, H.~R.~A., Pinfield, D.~J., et al.\ 2007, MNRAS, 374, 445

\bibitem[2006]{kilic06}
Kilic, M., Munn, J.~A., Harris, H.~C., et al.\ 2006, AJ, 131, 582

\bibitem[1999]{kirkpatrick99}
Kirkpatrick, J.~D., Reid, I.~N., Liebert, J., et al.\ 1999, ApJ, 519, 802

\bibitem[2005]{kirkpatrick05}
Kirkpatrick, J.~D.\ 2005, ARA\&A, 43, 195

\bibitem[2004]{knapp04}
Knapp, G.~R., Leggett, S.~K., Fan, X., et al.\ 2004, AJ, 127, 3553

\bibitem[2008]{lang08}
Lang, D., Hogg, D.~W., Jester, S., Rix, H.-W.\ 2008, in prep.

\bibitem[2007]{lawrence07}
Lawrence, A., Warren, S.~J., Almaini, O., et al.\ 2007, MNRAS, 379, 1599

\bibitem[1998]{leggett98}
Leggett, S.K., Ruiz, M.~T., Bergeron, P.\ 1998, ApJ, 497, 294

\bibitem[2003]{lepine03}
L{\'e}pine, S., Rich, R.~M., \& Shara, M.~M.\ 2003, ApJ, 591, L49

\bibitem[2004]{lepine04}
L{\'e}pine, S., Shara, M.~M., \& Rich, R.~M.\ 2004, ApJ, 602, L125

\bibitem[2005]{lepine05}
L{\'e}pine, S., Rich, R.~M., \& Shara, M.~M.\ 2005, ApJ, 633, L121

\bibitem[2005]{lepine05b}
L{\'e}pine, S., \& Shara, M.~M.\ 2005, AJ, 129, 1483

\bibitem[2007]{lepine07}
L{\'e}pine, S., Rich, R.~M., \& Shara, M.~M.\ 2007, ApJ, 669, 1235

\bibitem[2008]{lepine08}
L{\'e}pine, S., \& Scholz, R.-D.\ 2008, ApJ, 681, L33

\bibitem[2002]{liu02}
Liu, M.~C., Wainscoat, R.,
Mart{\'{\i}}n, E.~L., Barris, B., \& Tonry, J.\ 2002, ApJ, 568, L107

\bibitem[2007]{lodieu07}
Lodieu, N., Pinfield, D.~J., Leggett, S.~K., et al.\ 2007, MNRAS, 379, 1423

\bibitem[2007]{looper07}
Looper, D.~L., Kirkpatrick, J.~D., \& Burgasser, A.~J.\ 2007, AJ, 134, 1162

\bibitem[1999]{lupton99}
Lupton, R.~H., Gunn, J.~E., \& Szalay, A.~S.\ 1999, AJ, 118, 1406

\bibitem[2001]{lupton01}
Lupton, R., Gunn, J.~E., Ivezi{\'c}, Z., Knapp, G.~R., \& Kent, S.\ 2001,
in Astronomical Data Analysis Software and Systems X,
ed. F.~R. Harnden Jr., F.~A. Primini, H.~E. Payne (ASP, San Francisco),
ASP Conf. Ser., 238, 269

\bibitem[1922]{luyten22}
Luyten, W.~J.\ 1922, Lick Obs. Bull., 10, 135

\bibitem[1979a]{luyten79a}
Luyten, W.~J.\ 1979a, LHS Catalogue: a catalogue of stars with proper
motions exceeding 0.5" annually, University of Minnesota, Minneapolis

\bibitem[1979b]{luyten79b}
Luyten W.~J.\ 1979b, New Luyten Catalogue
     of Stars with Proper Motions Larger than Two Tenths of an Arcsecond,
     Univ. Minnesota, Minneapolis

\bibitem[1980]{luyten80}
Luyten W.~J., \& Hughes H.~S.\ 1980, NLTT first supplement,
     Univ. Minnesota, Minneapolis

\bibitem[1999]{martin99}
Mart\'{\i}n, E.~L., Delfosse, X., Basri, G., et al.\ 1999, AJ, 118, 2466

\bibitem[2006]{martin06}
Mart\'{\i}n, E.~L., Brandner, W., Bouy, H., Basri, G., Davis, J.,
Deshpande, R., \& Montgomery, M.~M.\ 2006, A\&A, 456, 253

\bibitem[2002]{mendez02}
M{\'e}ndez, R.~A.\ 2002, A\&A, 395, 779

\bibitem[2008]{metchev08}
Metchev, S.~A., Kirkpatrick, J.~D., Berriman, G.~B., \& Looper, D.\ 2008,
ApJ, 676, 1281

\bibitem[1992]{monet92}
Monet, D.~G., Dahn, C.~C., Vrba, F.~J., Harris, H.~C., Pier, J.~R., 
Luginbuhl, C.~B., \& Ables, H.~D.\ 1992, AJ, 103, 638

\bibitem[2000]{monet00}
Monet, D.~G., Fisher, M.~D., Liebert, J., et al.\ 2000, AJ, 120, 1541

\bibitem[2003]{monet03}
Monet, D.~G., Levine, S.~E., Canzian, B., et al.\ 2003, AJ, 125, 984

\bibitem[2004]{munn04}
Munn, J.~A., Monet, D.~G., Levine, S.~E., et al.\ 2004, AJ, 127, 3034

\bibitem[2001]{oppenheimer01}
Oppenheimer, B.~R., Hambly, N.~C., Digby, A.~P.,
     et al.\ 2001, Science, 292, 698

\bibitem[2008]{padmanabhan08}
Padmanabhan, N., Schlegel, D.~J., Finkbeiner, D.~P., et al.\ 2008,
ApJ, 674, 1217

\bibitem[2008]{phanbao08}
Phan-Bao, N., Bessell, M.~S., Mart\'{\i}n, E.~L., et al.\ 2008, MNRAS, 383, 831

\bibitem[2003]{pier03}
Pier, J.~R.; Munn, Jeffrey A., Hindsley, R.~B., et al.\ 2003, AJ, 125, 1559

\bibitem[2003]{pokorny03}
Pokorny, R.~S., Jones, H.~R.~A., \& Hambly, N.~C.\ 2003, A\&A, 397, 575

\bibitem[2004]{pokorny04}
Pokorny, R.~S., Jones, H.~R.~A., Hambly, N.~C., \& Pinfield, D.~J.\ 2004,
A\&A, 421, 763

\bibitem[2003]{raymond03}
Raymond, S.~N., Szkody, P., Hawley, S.~L., et al.\ 2003, AJ, 125, 2621

\bibitem[1995]{reid95}
Reid, I.~N., Hawley, S.~L., \& Gizis, J.~E.\ 1995, AJ, 110, 1838

\bibitem[2002]{reid02}
Reid, I.~N., Gizis, J.~E., \& Hawley, S.~L.\ 2002, AJ, 124, 2721

\bibitem[2001]{reid01}
Reid, I.~N., Sahu, K.~C., \& Hawley, S.~L.\ 2001, ApJ, 559, 942

\bibitem[2005]{reid05}
Reid, I.~N.\ 2005, ARA\&A, 43, 247

\bibitem[2008]{rowell08}
Rowell, N.~R., Kilic, M., \& Hambly, N.~C.\ 2008, MNRAS, 385, L23

\bibitem[1997]{ruiz97}
Ruiz, M.~T., Leggett, S.~K., \& Allard, F.\ 1997, ApJ, 491, L107

\bibitem[2004]{salim04}
Salim, S., Rich, R.~M., Hansen, B.~M., et al.\ 2004, ApJ, 601, 1075

\bibitem[2007]{schmidt07}
Schmidt, S.~J., Cruz, K.~L., Bongiorno, B.~J., Liebert, J., \& Reid, I.~N.\
2007, AJ, 133, 2258

\bibitem[2002]{schneider02}
Schneider, D.~P., Knapp, G.~R., Hawley, S.~L., et al.\ 2002, AJ, 123, 458

\bibitem[2000]{scholz00}
Scholz, R.-D., Irwin, M., Ibata, R., et al.\ 2000, A\&A, 353, 958

\bibitem[2002]{scholz02}
Scholz, R.-D., Szokoly, G.~P., Andersen, M., Ibata, R., \& Irwin, M.~J.\
2002, ApJ, 565, 539

\bibitem[2003]{scholz03}
Scholz, R.-D., McCaughrean, M.~J., Lodieu, N., \&
     Kuhlbrodt, B.\ 2003, A\&A, 398, L29

\bibitem[2004a]{scholz04a}
Scholz, R.-D., Lehmann, I., Matute, I., \& Zinnecker, H.\ 2004a, A\&A, 425, 519

\bibitem[2004b]{scholz04b}
Scholz, R.-D., McCaughrean, M.~J., \& Lodieu, N. 2004b, A\&A, 428, L25

\bibitem[2005]{scholz05}
Scholz, R.-D., Meusinger, H., Jahrei{\ss}, H.\ 2005, A\&A, 442, 211

\bibitem[2004]{sivarani04}
Sivarani, T., Kembhavi, A.~K., \& Gupchup, J.\ 2004, submitted to ApJ Letters
(this manuscript was previously available at:
http://www.iucaa.ernet.in/$\sim$gupchup/final\_ApjL.pdf)

\bibitem[2006]{skrutskie06}
Skrutskie, M.~F., Cutri, R.~M., Stiening, R., et al.\ 2006, AJ, 131, 1163

\bibitem[2002]{smith02}
Smith, J.~A., Tucker, D.~L., Kent, S., et al.\ 2002, AJ, 123, 2121

\bibitem[2007]{smith07}
Smith, M.~C., Ruchti, G.~R., Helmi, A., et al.\ 2007, MNRAS, 379, 755

\bibitem[2002]{stoughton02}
Stoughton, Ch., Lupton, R.~H., Bernardi, M., et al.\ 2002, AJ, 123, 485

\bibitem[2004]{subasavage04}
Subasavage, J.~P., Henry, T.~J., Hambly, N.~C.,
     Brown, M.~A., \& Jao, W.-Ch.\ 2004, AJ, 130, 1658

\bibitem[2008]{subasavage08}
Subasavage, J.~P., Henry, T.~J., Bergeron, P., Dufour, P.,
\& Hambly, N.~C.\ 2008, AJ, 136, 899

\bibitem[1998]{tinney98}
Tinney, C.~G.\ 1998, MNRAS, 296, L42

\bibitem[2008]{torres08}
Torres, S., Camacho, J., Isern, J., \& Garc{\'{\i}}a-Berro, E.\ 2008, 
A\&A, 486, 427

\bibitem[2006]{tucker06}
Tucker, D.~L., Kent, S., Richmond, M.~W., et al.\ 2006, AN, 327, 821

\bibitem[1995]{vanaltena95}
van Altena, W.~F., Lee, J.~T., \& Hoffleit, E.~D.\ 1995,
      The General Catalogue of Trigonometric Stellar
      Parallaxes, Fourth Edition, Yale University Observatory, available at
      http://vizier.u-strasbg.fr/viz-bin/Cat?I/238A

\bibitem[2004]{vrba04}
Vrba, F.~J., Henden, A.~A., Luginbuhl, C.~B., et al.\ 2004, AJ, 127, 2948

\bibitem[2007]{vidrih07}
Vidrih, S., Bramich, D.~M., Hewett, P.~C., et al.\ 2007, MNRAS, 382, 515

\bibitem[2007]{warren07}
Warren, S.~J., Mortlock, D.~J., Leggett, S.~K., et al.\ 2007, MNRAS, 381, 1400

\bibitem[2005]{west05}
West, A.~A., Walkowicz, L.~M., \& Hawley, S.~L.\ 2005, PASP, 117, 706

\bibitem[2008]{west08}
West, A.~A., Hawley, S.~L., Bochanski, J.~J., et al.\ 2008, 135, 785

\bibitem[2000]{york00}
York, D.~G., Adelman, J., Anderson, J.~E., et al.\ 2000, AJ, 120, 1579

\end{thebibliography}
\end{document}